\documentclass[aps,a4paper,superscriptaddress,preprintnumbers,showpacs,
amsmath,amssymb]{revtex4}
\usepackage{graphicx}
\usepackage{dcolumn}
\usepackage{color}
\usepackage{latexsym,amsfonts}
\usepackage{textcomp}
\usepackage{bm}

\usepackage{ulem}

\begin{document}

\title{Spin Precession of Slow Neutrons\\in Einstein--Cartan Gravity
  with Torsion, Chameleon and Magnetic Field}

\author{A. N. Ivanov}\email{ivanov@kph.tuwien.ac.at}
\affiliation{Atominstitut, Technische Universit\"at Wien, Stadionallee
  2, A-1020 Wien, Austria}
\author{M. Wellenzohn}\email{max.wellenzohn@gmail.com}
\affiliation{Atominstitut, Technische Universit\"at Wien, Stadionallee
  2, A-1020 Wien, Austria} \affiliation{FH Campus Wien, University of
  Applied Sciences, Favoritenstra\ss e 226, 1100 Wien, Austria}

\date{\today}

\begin{abstract}
We analyse a spin precession of slow neutrons in the Einstein--Cartan
gravity with torsion, chameleon and magnetic field. For the derivation
of the Heisenberg equation of motion of the neutron spin we use the
effective low--energy potential, derived by Ivanov and Wellenzohn
(Phys. Rev. D {\bf 92}, 125004 (2015)) for slow neutrons, coupled to
gravitational, chameleon and torsion fields to order $1/m$, where $m$
is the neutron mass. In addition to this low--energy interactions we
switch on the interaction of slow neutrons with a magnetic field. We
show that to linear order approximation with respect to gravitational,
chameleon and torsion fields the Dirac Hamilton operator for fermions
(neutrons), moving in spacetimes created by rotating coordinate
systems, contains the anti--Hermitian operators of torsion--fermion
(neutron) interactions, caused by torsion scalar and tensor
space--space--time and time--space--space degrees of freedom. Such
anti--Hermitian operators violate $CP$ and $T$ invariance. In the
low--energy approximation the $CP$ and $T$ violating torsion--fermion
(neutron) interactions appear only to order $O(1/m)$. One may assume
that in the rotating Universe and galaxies the obtained
anti--Hermitian torsion--fermion interactions might be an origin of i)
violation of $CP$ and $T$ invariance in the Universe and ii) of baryon
asymmetry. We show that anti--Hermitian
  torsion--fermion interactions of relativistic
    fermions, violating $CP$ and $T$ invariance, i) cannot be removed
  by non--unitary transformations of the Dirac fermion wave functions
  and ii) are conformal invariant. According to
  general requirements of conformal invariance of massive particle
  theories in gravitational fields (see R. H. Dicke, Phys. Rev. {\bf
    125}, 2163 (1962) and A. J. Silenko, Phys. Rev. {\bf D} 91, 065012
  (2015)), conformal invariance of anti--Hermitian torsion--fermion
  interactions is valid only if the fermion mass is changed by a
  conformal factor.
\end{abstract}

\maketitle

\section{Introduction}
\label{sec:introduction}

Recently \cite{Ivanov2015b} we have derived to order $O(1/m)$ the most
general effective low--energy potential for slow Dirac fermions with
mass $m$, coupled to gravitational, chameleon and torsion fields in
the Einstein--Cartan gravity. We have reduced the obtained potential
to order $O(1)$ in the large fermion mass expansion. We have shown
that the torsion pseudoscalar and tensor degrees of freedom can be, in
principle, measured in terrestrial laboratories through minimal
torsion--fermion couplings by using rotating devices. This is similar
to the experiments by Atwood {\it et al.}  \cite{Atwood1984} and by
Mashhoon \cite{Mashhoon1988}. These experiments used a rotating
two--crystal neutron interferometer and a neutron interferometer in a
rotating reference frame, respectively. We have assumed that the
measurements of the transition frequencies between quantum
gravitational states of ultracold neutrons in the qBounce experiments
\cite{Abele2010}--\cite{Jenke2014} as functions of an angular velocity
$\vec{\omega}$ of a rotating mirror should provide a new level of
highly precise probes of the properties of the Einstein--Cartan
gravity, dark energy and evolution of the Universe. In turn, the
measurements of the phase--shift of slow neutron wave function as a
function of an angular velocity $\vec{\omega}$ by a rotating neutron
interferometer \cite{Lemmel2015} should be of use for terrestrial
probes of new gravitational, chameleon and torsion interactions,
derived in \cite{Ivanov2015b}.

In this paper we propose an analysis of a spin precession of slow
neutrons in the Einstein--Cartan gravity with torsion, chameleon and
magnetic fields. As has been mentioned by Lehnert, Snow, and Yan
\cite{Lehnert2014}, a spin precession of slow neutrons is a very
sensitive technique to search for possible exotic neutron
interactions. In the experiment \cite{Lehnert2014} for the measurement
of an upper bound of the linear superposition of constant torsion
scalar and pseudoscalar degrees of freedom $\zeta$, caused by
torsion--fermion interactions by Kostelecky, Russell, and Tasson
\cite{Kostelecky2008}, a neutron spin rotation in the liquid ${^4}{\rm
  He}$ was investigated. The upper bound $|\zeta| < 9.1\times
10^{-23}\,{\rm GeV}$, obtained by Lehnert, Snow, and Yan
\cite{Lehnert2014}, is by a factor $10^5$ larger compared with the
estimate $|\zeta| < 10^{-27}\,{\rm GeV}$, obtained in
\cite{Ivanov2015a} by using the estimates in Table I by Kostelecky
     {\it et al.} \cite{Kostelecky2008}.  A spin dynamics of Dirac
     fermions with mass $m$ in curved spacetimes has been also
     investigated by Obukhov, Silenko, and Teryaev in
     \cite{Obukhov2009,Obukhov2011,Obukhov2014}.

The paper is organized as follows. In section \ref{sec:potential} we
calculate the effective low--energy potential, derived in
\cite{Ivanov2015b}, to linear order in gravitational, chameleon and
torsion fields in the spacetime with the Schwarzschild metric, taken
in the weak gravitational field approximation and modified by the
chameleon field and rotation with an angular velocity
$\vec{\omega}$. We show that the linearised effective low--energy
potential contains the anti--Hermitian interactions, vanishing at zero
angular velocity. In section \ref{sec:Hermiticity} we give a detailed
analysis of the anti--Hermitian interactions. In section
\ref{sec:equation} we derive the Heisenberg equation of motion for a
neutron spin precession in terms of the angular velocity operators,
caused by i) a magnetic field, ii) gravitational and chameleon fields
and iii) a torsion field, defining Hermitian and anti--Hermitian
torsion--fermion interactions. In section
  \ref{sec:obukhov} we show that the anti--Hermitian Hamilton operator
  and, correspondingly, the anti--Hermitian effective low--energy
  potential of torsion--fermion interactions, violating $CP$ and $T$
  invariance, i) cannot be removed by non--unitary (non--Hermitian)
  transformations \cite{Obukhov2014}--\cite{Gorbatenko2010} (see also
  \cite{Ivanov2015b}), ii) are conformal invariant
  \cite{Silenko2013,Silenko2015} and, referring to
    the experiments by Atwood {\it et al.}  \cite{Atwood1984} and by
    Mashhoon \cite{Mashhoon1988}, iii) can be in principle
  observable.  It should be emphasized that,
  according to the general requirements of conformal invariance of
  massive particle theories in gravitational fields under conformal
  transformation $\tilde{g}_{\mu\nu} \to O^2 \check{g}_{\mu\nu}$ (see,
  for example, Brans and Dicke \cite{Brans1961}, Dicke
  \cite{Dicke1962} and Silenko \cite{Silenko2015}), anti--Hermitian
  torsion--fermion interactions are conformal invariant only if the
  fermion mass $m$ is changed by a conformal factor $O$, i.e $m \to
  O^{-1}\check{m}$.  In section
  \ref{sec:conclusion} we summarize the obtained results and discuss
  some possible consequences of the anti--Hermitian torsion--fermion
  interactions. In the Appendix we give a detailed calculation of the
operator $G_{\hat{j}}$, responsible for the anti--Hermitian part of
Dirac Hamilton operator and of the effective low--energy potential of
slow fermions, coupled to gravitational, chameleon and torsion fields
to linear order approximation in curved spacetimes with rotation.

\section{Schr\"odinger--Pauli equation for slow neutrons in 
Einstein--Cartan gravity with torsion, chameleon and magnetic fields}
\label{sec:potential}

The Schr\"odinger--Pauli equation for slow neutrons, coupled to
gravitational, chameleon, torsion and magnetic fields in the
Einstein--Cartan gravity \cite{Ivanov2015b}, is equal to
\begin{eqnarray}\label{eq:1}
i\,\frac{\partial \Psi(t,\vec{r}\,)}{\partial t} = {\rm
  H}_{\rm FW}\,\Psi(t,\vec{r}\,)
\end{eqnarray}
where $\Psi(t,\vec{r}\,)$ is the wave function of slow neutrons and
${\rm H}$ is the Foldy--Wouthuysen Hamilton operator given by
\begin{eqnarray}\label{eq:2}
{\rm H}_{\rm FW} = - \frac{1}{2m}\,\Delta + m U_{\rm E} -
\vec{\mu}\cdot \vec{B} + \tilde{\Phi}_{\rm eff}(t,\vec{r},\vec{S}\,),
\end{eqnarray}
where $m$ is the neutron mass, $\Delta$ is the Laplace operator,
$U_{\rm E} = \vec{g}\cdot \vec{r}$~ is the gravitational potential of
the Earth with the Newtonian gravitational acceleration $\vec{g}$, and
$\vec{B}$ is a magnetic field.  The neutron magnetic dipole moment
$\vec{\mu} = 2\kappa_n\,\mu_N\,\vec{S}$ \cite{Sears1989} is expressed
in terms of the neutron anomalous magnetic moment $\kappa_n = -
1.9130427(5)$, measured in nuclear magnetons $\mu_N = e/2m_p =
3.1524512605(22)\times 10^{-8}\,{\rm eV\,T^{-1}}$ \cite{PDG2014},
which is defined in terms of the electric charge $e$ and mass $m_p$ of
the proton, and the neutron spin operator $\vec{S} =
\frac{1}{2}\,\vec{\sigma}$, where $\vec{\sigma}$ are the $2\times 2$
Pauli matrices \cite{Itzykson1980}.  Then, $\tilde{\Phi}_{\rm
  eff}(t,\vec{r},\vec{S}\,) = \Phi_{\rm eff}(t,\vec{r},\vec{S}\,) -
U_{\rm E}$, where $\Phi_{\rm eff}(t,\vec{r},\vec{S}\,)$ is the
effective low--energy potential for slow neutrons, coupled to
gravitational, chameleon and torsion fields (see Eq.(A.15) of
Ref.\cite{Ivanov2015b}). We would like to note that we have included
the interaction of slow neutrons with a magnetic field to linear order
approximation of the magnetic field. Since below we analyse the
contributions of the effective low--energy potential $\Phi_{\rm
  eff}(t,\vec{r}, \vec{S}\,)$ to linear order approximation of
gravitational, chameleon and torsion fields, we have neglected all
interactions, containing the products of a magnetic field with
gravitational, chameleon and torsion fields.

The calculation of the effective low-energy potential $\Phi_{\rm
  eff}(t, \vec{r},\vec{S}\,)$ we perform in the curved spacetime with
the line element \cite{Ivanov2015b} (see also \cite{Obukhov2011})
\begin{eqnarray}\label{eq:3}
d\tilde{s}^2 = \tilde{V}^2(x)\,dt^2 + \eta_{\hat{j}\hat{\ell}}
\tilde{W}^{\hat{j}}_j(x) \tilde{W}^{\hat{\ell}}_{\ell}(x) \Big(dx^j -
K^j(x)\, dt\Big)\Big(dx^{\ell} - K^{\ell}(x)\, dt\Big),
\end{eqnarray}
where $\hat{j} = 1,2,3$ and $j = 1,2,3$ are indices of the Minkowski
and curved spacetime, respectively, $\eta_{\hat{j}\hat{\ell}}$ is a
spatial part of the metric tensor in the Minkowski spacetime. The
functions $\tilde{V}^2(x)$ and $\tilde{W}^{\hat{j}}_j(x)$ are defined
by an arbitrary gravitational field. In comparison with Obukhov,
Silenko, and Teryaev \cite{Obukhov2011} the functions $\tilde{V}^2(x)$
and $\tilde{W}^{\hat{j}}_j(x)$ are modified by the chameleon field. In
turn, the functions $K^j(x)$, caused by rotations, are not modified by
the chameleon field. The vierbein fields in terms of which slow neutrons
couple to gravitational, chameleon and torsion fields in the
Einstein--Cartan gravity with the metric tensor in Eq.(\ref{eq:3}) are
equal to \cite{Ivanov2015b}
\begin{eqnarray}\label{eq:4}
\tilde{e}^{\hat{0}}_0(x) &=& \tilde{V}(x)\;,\;\tilde{e}^{\hat{j}}_0(x)
= - \tilde{W}^{\hat{j}}_j(x)K^j(x)\;,\;\tilde{e}^{\hat{0}}_j(x) =
0\;,\;\tilde{e}^{\hat{j}}_j(x) =
\tilde{W}^{\hat{j}}_j(x),\nonumber\\ \tilde{e}^0_{\hat{0}}(x) &=&
\frac{1}{\tilde{V}(x)}\;,\; \tilde{e}^0_{\hat{j}}(x) = 0\;,\;
\tilde{e}^j_{\hat{0}}(x) = \frac{K^j(x)}{\tilde{V}(x)}\;,\;
\tilde{e}^j_{\hat{j}}(x) = \tilde{W}^j_{\hat{j}}(x).
\end{eqnarray}
The vierbein fields in Eq.(\ref{eq:4}) have been calculated at the
assumption that the functions $\tilde{W}^{\hat{j}}_j(x)$ and
$\tilde{W}^j_{\hat{j}}(x)$ obey the orthogonality relations
\cite{Ivanov2015b}
\begin{eqnarray}\label{eq:5}
\tilde{W}^{\hat{j}}_j(x) \tilde{W}^j_{\hat{\ell}} =
\delta^{\hat{j}}_{\hat{\ell}}\;,\; \tilde{W}^j_{\hat{j}}(x)
\tilde{W}^{\hat{j}}_{\ell} = \delta^j_{\ell},
\end{eqnarray}
which are fulfilled for the Schwarzschild metric in the weak
gravitational field approximation \cite{Ivanov2015a}. In terms of the
vierbein fields Eq.(\ref{eq:4}) the effective low--energy potential
$\Phi_{\rm eff}(t, \vec{r}, \vec{S}\,)$ is given by
\begin{eqnarray}\label{eq:6}
\hspace{-0.3in}&&\Phi_{\rm eff}(t, \vec{r},\vec{S}\,) = (A - 1)\,m + B
+ 2C^{\hat{\ell}}S_{\hat{\ell}} + i\,L^j\,\frac{\partial}{\partial
  x^j} + \frac{1}{2 m
}\,\eta^{\hat{j}\hat{k}}\,\Big(\frac{D^j_{\hat{j}}D^k_{\hat{k}}}{A} -
\delta^j_{\hat{j}} \delta^k_{\hat{k}}\Big)\,
\frac{\partial^2}{\partial x^j \partial x^k}\nonumber\\
\hspace{-0.3in}&& + \frac{1}{4 m
  A}\,\eta^{\hat{j}\hat{k}}\,D^j_{\hat{j}}\frac{\partial
  D^k_{\hat{k}}}{\partial x^j}\,\frac{\partial}{\partial x^k} +
\frac{1}{4 m}\,\eta^{\hat{j}\hat{k}}\,D^k_{\hat{k}}\,\frac{\partial
}{\partial
  x^k}\Big(\frac{D^j_{\hat{j}}}{A}\Big)\,\frac{\partial}{\partial x^j}
+ \frac{1}{2 m A}\,i\,\epsilon^{\hat{j}\hat{k}\hat{\ell}}\,
S_{\hat{\ell}}\,D^j_{\hat{j}}\frac{\partial D^k_{\hat{k}}}{\partial
  x^j}\,\frac{\partial}{\partial x^k}\nonumber\\ 
\hspace{-0.3in}&& -
\frac{1}{2 m}\,i\,\epsilon^{\hat{j}\hat{k}\hat{\ell}}\,
S_{\hat{\ell}}\,D^k_{\hat{k}}\, \frac{\partial }{\partial
  x^k}\Big(\frac{D^j_{\hat{j}}}{A}\Big)\,\frac{\partial}{\partial x^j}
+ \frac{1}{2 m A}\,\eta^{\hat{j}\hat{k}}\,G_{\hat{j}}\,D^k_{\hat{k}}\,
\frac{\partial}{\partial x^k} + \frac{1}{4 m
}\,\eta^{\hat{j}\hat{k}}\,D^k_{\hat{k}}\, \frac{\partial}{\partial
  x^k}\Big(\frac{G_{\hat{j}}}{A}\Big)\nonumber\\ \hspace{-0.3in}&& -
\frac{1}{2 m }\,i\,\epsilon^{\hat{j}\hat{k}\hat{\ell}}\,
S_{\hat{\ell}}\,D^k_{\hat{k}}\,\frac{\partial}{\partial
  x^k}\Big(\frac{G_{\hat{j}}}{A}\Big)+ \frac{1}{m
  A}\,i\,S^{\hat{k}}\,K\, D^k_{\hat{k}}\,\frac{\partial }{\partial
  x^k} + \frac{1}{2 m}\,i\,S^{\hat{k}}\,D^k_{\hat{k}}\,
\frac{\partial}{\partial x^k}\Big(\frac{K}{A}\Big) + \frac{1}{2 m
  A}\,\eta^{\hat{j}\hat{k}}\,G_{\hat{k}}\,D^j_{\hat{j}}
\frac{\partial}{\partial x^j}\nonumber\\ 
\hspace{-0.3in}&&+ \frac{1}{4 m
  A}\,\eta^{\hat{j}\hat{k}}\,D^j_{\hat{j}}\frac{\partial
  G_{\hat{k}}}{\partial x^j} + \frac{1}{2 m
  A}\,i\,\epsilon^{\hat{j}\hat{k}\hat{\ell}}\,
S_{\hat{\ell}}\,D^j_{\hat{j}}\frac{\partial G_{\hat{k}}}{\partial x^j}
+ \frac{1}{2 m A}\,\eta^{\hat{j}\hat{k}}\,G_{\hat{j}}\,G_{\hat{k}} +
\frac{1}{m A}\,i\,S^{\hat{k}} K\,G_{\hat{k}} + \frac{1}{m
  A}\,i\,S^{\hat{j}}\,K\,D^j_{\hat{j}}\, \frac{\partial}{\partial
  x^j}\nonumber\\
\hspace{-0.3in}&& + \frac{1}{2 m A}\,i\,S^{\hat{j}}\,D^j_{\hat{j}}\,
\frac{\partial K}{\partial x^j} + \frac{1}{m
  A}\,i\,S^{\hat{j}}\,K\,G_{\hat{j}} + \frac{1}{2 m A}\,K^2 -
\frac{1}{4 m
  A^2}\,\eta^{\hat{j}\hat{k}}\,D^j_{\hat{j}}\,D^k_{\hat{k}}\,\frac{\partial
  A}{\partial x^k}\, \frac{\partial}{\partial x^j} - \frac{1}{8 m
  A}\,\eta^{\hat{j}\hat{k}}\,D^j_{\hat{j}}\, \frac{\partial}{\partial
  x^j}\Big(\frac{D^k_{\hat{k}}}{A}\,\frac{\partial A}{\partial
  x^k}\Big)\nonumber\\
\hspace{-0.3in}&& - \frac{1}{4 m
  A}\,i\,\epsilon^{\hat{j}\hat{k}\hat{\ell}}\,
S_{\hat{\ell}}\,D^j_{\hat{j}}\, \frac{\partial}{\partial
  x^j}\Big(\frac{D^k_{\hat{k}}}{A}\,\frac{\partial A}{\partial
  x^k}\Big) - \frac{1}{4 m A^2}\,\eta^{\hat{j}\hat{k}}\,
G_{\hat{j}}\,D^k_{\hat{k}}\,\frac{\partial A}{\partial x^k} -
\frac{1}{2 m A^2}\,i\,S^{\hat{k}}\, K\,D^k_{\hat{k}}\,\frac{\partial
  A}{\partial x^k} +
\frac{1}{4mA^3}\,\eta^{\hat{j}\hat{k}}\,D^j_{\hat{j}}
D^k_{\hat{k}}\,\frac{\partial A}{\partial x^j}\, \frac{\partial
  A}{\partial x^k}.
\end{eqnarray}
The operators $A$, $B$, $C^{\hat{\ell}}$, $D^j_{\hat{j}}$, $G_{\hat{j}}$, $K$ and $L^j$ are equal to \cite{Ivanov2015b}
\begin{eqnarray}\label{eq:7}
A &=& \tilde{e}^{\hat{0}}_0(x),\nonumber\\ B &=& -
\frac{1}{2}\,i\,\tilde{e}^{\hat{0}}_0(x)\,\frac{1}{\sqrt{-
    \tilde{g}}}\,\frac{\partial }{\partial x^j}\Big(\sqrt{-
  \tilde{g}}\,\tilde{e}^j_{\hat{0}}(x)\Big) +
\frac{1}{2}\,i\,(\tilde{e}^{\hat{0}}_0(x))^2\,\tilde{e}^j_{\hat{0}}(x)
\,\frac{1}{\sqrt{- \tilde{g}(x)}}\frac{\partial }{\partial
  x^j}\Big(\sqrt{-
  \tilde{g}(x)}\,\tilde{e}^0_{\hat{0}}(x)\Big),\nonumber\\ C^{\hat{\ell}}
&=& \frac{1}{4}\,\tilde{e}^{\hat{0}}_0(x)\,\Big(\tilde{\omega}_{0
  \hat{j}\hat{k}}(x)\,\tilde{e}^0_{\hat{0}}(x) + \tilde{\omega}_{\ell
  \hat{j}\hat{k}}(x)\,\tilde{e}^{\ell}_{\hat{0}}(x) +
\tilde{\omega}_{\ell \hat{0}\hat{j}}(x)\,
\tilde{e}^{\ell}_{\hat{k}}(x) - \tilde{\omega}_{\ell \hat{j}
  \hat{0}}(x)\, \tilde{e}^{\ell}_{\hat{k}}(x)\Big)\,
\epsilon^{\hat{j}\hat{k}\hat{\ell}},\nonumber\\ D^j_{\hat{j}} &=& -
\tilde{e}^{\hat{0}}_0(x)\,\tilde{e}^j_{\hat{j}}(x),\nonumber\\ G_{\hat{j}}
&=& \frac{1}{2}\, \tilde{e}^{\hat{0}}_0(x)\,\Big({\tilde{\cal
    T}^{\alpha}\,\!\!}_{\alpha\ell}(x) \,\tilde{e}^{\ell}_{\hat{j}}(x)
+ \tilde{\omega}_{0\hat{j}\hat{0}}(x)\, \tilde{e}^0_{\hat{0}}(x) +
\tilde{\omega}_{\ell\hat{j}\hat{0}}(x)\, \tilde{e}^{\ell}_{\hat{0}}(x)
+ \tilde{\omega}_{\ell\hat{j}\hat{k}}(x)\,
\tilde{e}^{\ell}_{\hat{\ell}}(x)\,\eta^{\hat{\ell}\hat{k}}\Big)\nonumber\\ &+&
\frac{1}{2}\,(\tilde{e}^{\hat{0}}_0(x))^2\,\tilde{e}^j_{\hat{j}}(x)
\,\frac{1}{\sqrt{- \tilde{g}(x)}}\,\frac{\partial }{\partial
  x^j}\Big(\sqrt{-
  \tilde{g}(x)}\,\tilde{e}^0_{\hat{0}}(x)\Big),\nonumber\\ K &=& -
\frac{1}{4}\,\tilde{\omega}_{\ell
  \hat{j}\hat{k}}(x)\,\tilde{e}^{\hat{0}}_0(x)\,
\tilde{e}^{\ell}_{\hat{\ell}}(x)\,\epsilon^{\hat{j}\hat{k}\hat{\ell}},\nonumber\\ L^j
&=& - \tilde{e}^{\hat{0}}_0(x)\,\tilde{e}^j_{\hat{0}}(x).
\end{eqnarray}
Here $\tilde{g}(x) = - {\rm det}\{\tilde{g}_{\mu\nu}(x)\}$, where the
metric tensor in the Jordan frame $\tilde{g}_{\mu\nu}(x)$ is related
to the metric tensor in the Einstein frame $g_{\mu\nu}(x)$ by
$\tilde{g}_{\mu\nu}(x) = f^2(x)\,g_{\mu\nu}(x)$
\cite{Chameleon1,Chameleon2}, where the conformal factor $f(x) =
e^{\,\beta\,\phi(x)/M_{\rm Pl}}$ is defined in terms of the chameleon
field $\phi(x)$, the chameleon--matter coupling constant $\beta$ and
the reduced Planck mass $M_{\rm Pl} = 1/\sqrt{8\pi G_N} = 2.435\times
10^{27}\,{\rm eV}$ with the Newtonian gravitational constant $G_N$
\cite{PDG2014}. The spin connection
$\tilde{\omega}_{\mu\hat{\alpha}\hat{\beta}}(x)$ is defined by
\cite{Ivanov2015a,Ivanov2015b}
\begin{eqnarray}\label{eq:8}
\tilde{\omega}_{\mu\hat{\alpha}\hat{\beta}}(x) = -
\eta_{\hat{\alpha}\hat{\varphi}}\Big(\partial_{\mu}\tilde{e}^{\hat{\varphi}}_{\nu}(x)
- {\tilde{\Gamma}^{\alpha}\,}_{\mu\nu}(x)
\tilde{e}^{\hat{\varphi}}_{\alpha}(x)\Big)\tilde{e}^{\nu}_{\hat{\beta}}(x).
\end{eqnarray}
where $\alpha = 0,1,2,3$ and $\hat{\alpha} = 0,1,2,3$ are the indices
in the 4--dimensional curved and Minkowski spacetime,
respectively. The affine connection
${\tilde{\Gamma}^{\alpha}\,}_{\mu\nu}(x)$ is determined by
\begin{eqnarray}\label{eq:9}
{\tilde{\Gamma}^{\alpha}\,}_{\mu\nu}(x) =
\widetilde{\{{^\alpha}_{\mu\nu}\}} + {\tilde{\cal
    K}^{\alpha}\,}_{\mu\nu}(x),
\end{eqnarray}
where $\widetilde{\{{^\alpha}_{\mu\nu}\}}$ are the Christoffel symbols
\cite{LL2008}
\begin{eqnarray}\label{eq:10}
\widetilde{\{{^\alpha}_{\mu\nu}\}} = \frac{1}{2}\,\tilde{g}^{\alpha\lambda}\Big(\frac{\partial
  \tilde{g}_{\lambda\mu}}{\partial x^{\nu}} + \frac{\partial
  \tilde{g}_{\lambda\nu}}{\partial x^{\mu}} - \frac{\partial
  \tilde{g}_{\mu\nu}}{\partial x^{\lambda}}\Big)
\end{eqnarray}
and ${\tilde{\cal K}^{\alpha}\,}_{\mu\nu}(x) = - \frac{1}{2}(
{\tilde{\cal T}^{\alpha}\,}_{\mu\nu}(x) - {{\tilde{\cal
      T}_{\mu}\,}^{\alpha}\,}_{\nu}(x) - {{\tilde{\cal
      T}_{\nu}\,}^{\alpha}\,}_{\mu}(x))$ is the contorsion tensor,
expressed in terms of the torsion field ${\tilde{\cal
    T}^{\alpha}\,}_{\mu\nu}(x) =
\tilde{g}^{\alpha\sigma}(x)\,\tilde{\cal T}_{\sigma\mu\nu}(x)$
\cite{Ivanov2015a}. For the analysis of the effective potential
Eq.(\ref{eq:6}) we assume a motion of Dirac fermions with mass $m$ in
the curved spacetime with the Schwarzschild metric, taken in the weak
gravitational field approximation and modified by the contributions of
the chameleon field and rotation. The line element of such a spacetime
is given by \cite{Ivanov2015b}
\begin{eqnarray}\label{eq:11}
d\tilde{s}^2 = (1 + 2 U_+)\,dt^2 + 2\,(1 - 2 U_-)\,\vec{K}\cdot
d\vec{r}\,dt - (1 - 2 U_-)\,d\vec{r}^{\,2},
\end{eqnarray}
where we have neglected the contribution of the terms of order
$\vec{K}^{\,2}$ that is well justified in terrestrial laboratories
\cite{Hehl1990} and kept the contributions of the chameleon field to
linear order. The potentials $U_{\pm}$ are equal to \cite{Ivanov2015a}
\begin{eqnarray}\label{eq:12}
U_{\pm} = U_{\rm E} \pm \frac{\beta}{M_{\rm Pl}}\,\phi(x).
\end{eqnarray}
To linear order contributions of the gravitational and chameleon field
the vierbein fields Eq.(\ref{eq:4}) read
\begin{eqnarray}\label{eq:13}
\tilde{e}^{\hat{0}}_0(x) &=&1 + U_+\;,\; \tilde{e}^{\hat{j}}_0(x) = -
(1 - U_-)\,K^{\hat{j}}(x)\;,\; \tilde{e}^{\hat{0}}_j(x) = 0\;,\;
\tilde{e}^{\hat{j}}_j(x) = (1 - U_-)\,
\delta^{\hat{j}}_j,\nonumber\\ \tilde{e}^0_{\hat{0}}(x) &=& 1 -
U_+\;,\; \tilde{e}^j_{\hat{0}}(x) = +(1 - U_+)\, K^j(x)\;,\;
\tilde{e}^0_{\hat{j}}(x) = 0\;,\; \tilde{e}^j_{\hat{j}}(x) = (1 +
U_-)\,\delta^j_{\hat{j}}.
\end{eqnarray}
 In the spacetime with metric Eq.(\ref{eq:11}) and the vierbein fields
 Eq.(\ref{eq:13}) the operators $A$, $B$, $C^{\hat{\ell}}$,
 $D^j_{\hat{j}}$, $G_{\hat{j}}$, $K$ and $L^j$, calculated to linear
 approximation in gravitational, chameleon and torsion fields, are
 equal to
\begin{eqnarray}\label{eq:14}
A &=& 1 + U_+,\nonumber\\ B &=& - \frac{1}{2}\,i\,{\rm
  div}\,\vec{K},\nonumber\\ C^{\hat{\ell}} &=& - \frac{1}{4}\,({\rm
  rot}\,\vec{K}\,)^{\hat{\ell}} +
\frac{1}{8}\,\epsilon^{\hat{\ell}\hat{j}\hat{k}}\,({\cal
  T}_{\hat{j}\hat{k}\hat{0}} + {\cal T}_{\hat{k}\hat{0}\hat{j}} +
     {\cal T}_{\hat{0}\hat{j}\hat{k}}) +
     \frac{1}{4}\,\epsilon^{\hat{\ell} \hat{j}
       \hat{k}}\,K_{\hat{j}}\,{\cal T}_{\hat{0}\hat{0}\hat{k}} +
     \frac{1}{4}\,\epsilon^{\hat{\ell} \hat{j} \hat{k}}\,{\cal
       T}_{\hat{j}\hat{k}\hat{a}}\,K^{\hat{a}} = \nonumber\\ &=& -
     \frac{1}{4}\,({\rm rot}\,\vec{K}\,)^{\hat{\ell}} +
     \frac{1}{4}\,{\cal B}^{\hat{\ell}} + \frac{1}{6}\,{\cal
       K}\,K^{\hat{\ell}} + \frac{1}{4}\,\epsilon^{\hat{\ell} \hat{j}
       \hat{k}}\,K_{\hat{j}}\,{\cal M}_{\hat{0}\hat{0}\hat{k}} +
     \frac{1}{4}\,\epsilon^{\hat{\ell} \hat{j} \hat{k}}\,{\cal
       M}_{\hat{j}\hat{k}\hat{a}}\,K^{\hat{a}},\nonumber\\ D^j_{\hat{j}}
     &=& - (1 + U_++ U_-)\,\delta^j_{\hat{j}},\nonumber\\ G_{\hat{j}}
     &=& - \frac{1}{2}\,\frac{\partial}{\partial x^{\hat{j}}}(U_+ +
     U_-) + \frac{1}{2}\,({\cal T}_{\hat{0}\hat{j}\hat{\ell}} + {\cal
       T}_{\hat{\ell}\hat{j}\hat{0}}) K^{\hat{\ell}} +
     \frac{1}{2}\,K_{\hat{j}}\,{\cal
       T}_{\hat{\ell}\hat{k}\hat{0}}\,\eta^{\hat{\ell}\hat{k}}
     =\nonumber\\ &=& - \frac{1}{2}\,\frac{\partial}{\partial
       x^{\hat{j}}}(U_+ + U_-) + \frac{2}{3}\,{\cal
       E}_{\hat{0}}\,K_{\hat{j}} + \frac{1}{2}\,({\cal
       M}_{\hat{\ell}\hat{j}\hat{0}} + {\cal
       M}_{\hat{0}\hat{j}\hat{\ell}} )\,K^{\hat{\ell}},\nonumber\\ K
     &=& - \frac{1}{8}\,\epsilon^{\hat{j}\hat{k}\hat{\ell}}\,{\cal
       T}_{\hat{j}\hat{k}\hat{\ell}} +
     \frac{1}{8}\,\epsilon^{\hat{j}\hat{k}\hat{\ell}}\,K_{\hat{j}}\,{\cal
       T}_{\hat{0}\hat{k}\hat{\ell}} =\nonumber\\ &=& -
     \frac{1}{4}\,{\cal K} + \frac{1}{12}\,\vec{K}\cdot \vec{\cal B} +
     \frac{1}{8}\,\epsilon^{\hat{j}\hat{k}\hat{\ell}}\,K_{\hat{j}}\,{\cal
       M}_{\hat{0}\hat{k}\hat{\ell}}, \nonumber\\ L^j &=& - K^j,
\end{eqnarray}
where ${\cal B}^{\hat{\ell}} =
\frac{1}{2}\,\epsilon^{\hat{\ell}\hat{j}\hat{k}}\,({\cal
  T}_{\hat{j}\hat{k}\hat{0}} + {\cal T}_{\hat{k}\hat{0}\hat{j}} +
     {\cal T}_{\hat{0}\hat{j}\hat{k}})$, ${\cal K} =
     \frac{1}{2}\,\epsilon^{\hat{j}\hat{k}\hat{\ell}}\,{\cal
       T}_{\hat{j}\hat{k}\hat{\ell}}$, ${\cal
       M}_{\hat{\sigma}\hat{\mu}\hat{\nu}}$ (${\cal
       M}_{\hat{0}\hat{0}\hat{k}}$, ${\cal
       M}_{\hat{0}\hat{j}\hat{\ell}}$, ${\cal
       M}_{\hat{\ell}\hat{j}\hat{0}}$ and ${\cal
       M}_{\hat{j}\hat{k}\hat{\ell}}$) and ${\cal E}_{\hat{0}} =
          {{\cal T}^{\hat{\alpha}}}_{\hat{\alpha}\hat{0}}$ are torsion
          axial--vector, pseudoscalar, tensor and scalar degrees of
          freedom, respectively \cite{Ivanov2015a,Ivanov2015b}. For
          the calculation of the operators in Eq.(\ref{eq:14}) we have
          used the following irreducible representation of the torsion
          field ${\cal T}_{\sigma\mu\nu}$ \cite{Kostelecky2008} (see
          also \cite{Ivanov2015a,Ivanov2015b})
\begin{eqnarray}\label{eq:15}
{\cal T}_{\hat{\sigma}\hat{\mu}\hat{\nu}}(x) =
\frac{1}{3}\,\Big(\eta_{\hat{\sigma}\hat{\mu}} {\cal E}_{\hat{\nu}}(x) -
\eta_{\hat{\sigma}\hat{\nu}} {\cal E}_{\hat{\mu}}(x)\Big) +
\frac{1}{3}\,\epsilon_{\hat{\sigma}\hat{\mu}\hat{\nu}\hat{\alpha}}\, {\cal B}^{\hat{\alpha}}(x) +
     {\cal M}_{\hat{\sigma}\hat{\mu}\hat{\nu}}(x).
\end{eqnarray}
The torsion field ${\cal T}_{\hat{\sigma}\hat{\mu}\hat{\nu}}(x)$,
antisymmetric with respect to indices $\hat{\mu}$ and $\hat{\nu}$,
possesses 24 independent components, where the 4--vector ${\cal
  E}_{\hat{\nu}}(x)$ and axial 4--vector ${\cal B}^{\hat{\alpha}}(x)$
fields with 4 independent degrees of freedom each are defined by
\begin{eqnarray}\label{eq:16}
{\cal E}_{\hat{\nu}}(x) = \eta^{\hat{\sigma}\hat{\mu}}\,{\cal
  T}_{\hat{\sigma}\hat{\mu}\hat{\nu}}(x)\quad,\quad {\cal
  B}^{\hat{\alpha}}(x) =
\frac{1}{2}\,\epsilon^{\hat{\alpha}\hat{\sigma}\hat{\mu}\hat{\nu}}\,{\cal
  T}_{\hat{\sigma}\hat{\mu}\hat{\nu}}(x).
\end{eqnarray}
The residual 16 degrees of freedom are absorbed by the tensor ${\cal
  M}_{\hat{\alpha}\hat{\mu}\hat{\nu}}$, which obeys the constraints
$\eta^{\hat{\sigma}\hat{\mu}}{\cal M}_{\hat{\sigma}\hat{\mu}\hat{\nu}}
= \epsilon^{\hat{\alpha}\hat{\sigma}\hat{\mu}\hat{\nu}}{\cal
  M}_{\hat{\sigma}\hat{\mu}\hat{\nu}} = 0$. Then,
$\epsilon_{\hat{\sigma}\hat{\mu}\hat{\nu}\hat{\alpha}}$ and
$\epsilon^{\hat{\alpha}\hat{\sigma}\hat{\mu}\hat{\nu}}$ are the
Levi--Civita tensors such as $\epsilon_{\hat{0}\hat{1}\hat{2}\hat{3}}
= - \epsilon^{\hat{0}\hat{1}\hat{2}\hat{3}} = - 1$
\cite{Itzykson1980}.  For the derivation of the axial--vector field
${\cal B}^{\hat{\alpha}}$ in terms of the torsion field ${\cal
  T}_{\hat{\sigma}\hat{\mu}\hat{\nu}}(x)$ we have used the relation
$\epsilon^{\hat{\alpha}\hat{\sigma}\hat{\mu}\hat{\nu}}
\epsilon_{\hat{\sigma}\hat{\mu}\hat{\nu}\hat{\beta}} = -
6\,\delta^{\hat{\alpha}}_{\hat{\beta}}$ \cite{Itzykson1980}. Now we
rewrite the effective low--energy potential Eq.(\ref{eq:6}) omitting
the terms, which contributions are  smaller compared to the terms of the
linear order approximation
\begin{eqnarray}\label{eq:17}
\hspace{-0.3in}&&\Phi_{\rm eff}(t, \vec{r},\vec{S}\,) = (A - 1)\,m + B
+ 2C^{\hat{\ell}}S_{\hat{\ell}} + i\,L^j\,\frac{\partial}{\partial
  x^j} + \frac{1}{2 m
}\,\eta^{\hat{j}\hat{k}}\,\Big(\frac{D^j_{\hat{j}}D^k_{\hat{k}}}{A} -
\delta^j_{\hat{j}} \delta^k_{\hat{k}}\Big)\,
\frac{\partial^2}{\partial x^j \partial x^k}\nonumber\\
\hspace{-0.3in}&& + \frac{1}{4 m
  A}\,\eta^{\hat{j}\hat{k}}\,D^j_{\hat{j}}\frac{\partial
  D^k_{\hat{k}}}{\partial x^j}\,\frac{\partial}{\partial x^k} +
\frac{1}{4 m}\,\eta^{\hat{j}\hat{k}}\,D^k_{\hat{k}}\,\frac{\partial
}{\partial
  x^k}\Big(\frac{D^j_{\hat{j}}}{A}\Big)\,\frac{\partial}{\partial x^j}
+ \frac{1}{2 m A}\,i\,\epsilon^{\hat{j}\hat{k}\hat{\ell}}\,
S_{\hat{\ell}}\,D^j_{\hat{j}}\frac{\partial D^k_{\hat{k}}}{\partial
  x^j}\,\frac{\partial}{\partial x^k}\nonumber\\  
\hspace{-0.3in}&& - \frac{1}{2
  m}\,i\,\epsilon^{\hat{j}\hat{k}\hat{\ell}}\,
S_{\hat{\ell}}\,D^k_{\hat{k}}\, \frac{\partial }{\partial
  x^k}\Big(\frac{D^j_{\hat{j}}}{A}\Big)\,\frac{\partial}{\partial x^j}
+ \frac{1}{2 m A}\,\eta^{\hat{j}\hat{k}}\,G_{\hat{j}}\,D^k_{\hat{k}}\,
\frac{\partial}{\partial x^k} + \frac{1}{4 m
}\,\eta^{\hat{j}\hat{k}}\,D^k_{\hat{k}}\, \frac{\partial}{\partial
  x^k}\Big(\frac{G_{\hat{j}}}{A}\Big)\nonumber\\ 
\hspace{-0.3in}&& -
\frac{1}{2 m }\,i\,\epsilon^{\hat{j}\hat{k}\hat{\ell}}\,
S_{\hat{\ell}}\,D^k_{\hat{k}}\,\frac{\partial}{\partial
  x^k}\Big(\frac{G_{\hat{j}}}{A}\Big)+ \frac{1}{m
  A}\,i\,S^{\hat{k}}\,K\, D^k_{\hat{k}}\,\frac{\partial }{\partial
  x^k} + \frac{1}{2 m}\,i\,S^{\hat{k}}\,D^k_{\hat{k}}\,
\frac{\partial}{\partial x^k}\Big(\frac{K}{A}\Big) + \frac{1}{2 m
  A}\,\eta^{\hat{j}\hat{k}}\,G_{\hat{k}}\,D^j_{\hat{j}}
\frac{\partial}{\partial x^j}\nonumber\\
\hspace{-0.3in}&&+ \frac{1}{4 m
  A}\,\eta^{\hat{j}\hat{k}}\,D^j_{\hat{j}}\frac{\partial
  G_{\hat{k}}}{\partial x^j} + \frac{1}{2 m
  A}\,i\,\epsilon^{\hat{j}\hat{k}\hat{\ell}}\,
S_{\hat{\ell}}\,D^j_{\hat{j}}\frac{\partial G_{\hat{k}}}{\partial x^j}
+ \frac{1}{m A}\,i\,S^{\hat{j}}\,K\,D^j_{\hat{j}}\,
\frac{\partial}{\partial x^j} + \frac{1}{2 m
  A}\,i\,S^{\hat{j}}\,D^j_{\hat{j}}\, \frac{\partial K}{\partial x^j}
\nonumber\\ 
\hspace{-0.3in}&& - \frac{1}{4 m
  A^2}\,\eta^{\hat{j}\hat{k}}\,D^j_{\hat{j}}\,D^k_{\hat{k}}\,\frac{\partial
  A}{\partial x^k}\, \frac{\partial}{\partial x^j} - \frac{1}{8 m
  A}\,\eta^{\hat{j}\hat{k}}\,D^j_{\hat{j}}\, \frac{\partial}{\partial
  x^j}\Big(\frac{D^k_{\hat{k}}}{A}\,\frac{\partial A}{\partial
  x^k}\Big) - \frac{1}{4 m
  A}\,i\,\epsilon^{\hat{j}\hat{k}\hat{\ell}}\,
S_{\hat{\ell}}\,D^j_{\hat{j}}\, \frac{\partial}{\partial
  x^j}\Big(\frac{D^k_{\hat{k}}}{A}\,\frac{\partial A}{\partial
  x^k}\Big).
\end{eqnarray}
Plugging the operators Eq.(\ref{eq:14}) into Eq.(\ref{eq:17}) we
arrive at the following effective low--energy potential
\begin{eqnarray}\label{eq:18}
\tilde{\Phi}_{\rm eff}(t,\vec{r},\vec{S}\,) &=& \Phi^{(1)}_{\rm
  eff}(t,\vec{r},\vec{S}\,) + \Phi^{(2)}_{\rm
  eff}(t,\vec{r},\vec{S}\,) + \Phi^{(3)}_{\rm
  eff}(t,\vec{r},\vec{S}\,) + \Phi^{(4)}_{\rm
  eff}(t,\vec{r},\vec{S}\,),
\end{eqnarray}
where we have denoted:
\begin{eqnarray}\label{eq:19}
\Phi^{(1)}_{\rm eff}(t,\vec{r},\vec{S}\,) &=& m\,(U_+ - U_{\rm E}) -
i\,\vec{K}\cdot \vec{\nabla} - i\,\frac{1}{2}\,{\rm div}\vec{K} +
\frac{1}{2}\, \vec{S}\cdot {\rm rot}\vec{K} -
\frac{1}{2}\,\vec{S}\cdot \vec{\cal B}\nonumber\\ &-&
\frac{1}{3}\,{\cal K}\,\vec{S}\cdot \vec{K} -
\frac{1}{2}\,\vec{S}\cdot (\vec{K}\times \vec{\cal M}\,) +
\frac{1}{2}\,S_{\hat{j}}\,\epsilon^{\hat{j}\hat{k}\hat{\ell}}\,{\cal
  M}_{\hat{k}\hat{\ell}\hat{a}}\,K^{\hat{a}}
\end{eqnarray}
with $(\vec{\cal M}\,)_{\hat{k}} = - {\cal
  M}_{\hat{0}\hat{0}\hat{k}}$. The effective low--energy potential
Eq.(\ref{eq:19}) agrees well with the result, obtained in
\cite{Ivanov2015b}. Then, the potential $\Phi^{(2)}_{\rm
  eff}(t,\vec{r},\vec{S}\,)$, taking the form
\begin{eqnarray}\label{eq:20}
\Phi^{(2)}_{\rm eff}(t,\vec{r},\vec{S}\,) &=& -
\frac{1}{2m}\,\vec{\nabla}(U_+ + 2 U_-)\cdot \vec{\nabla} -
\frac{1}{2m}\,(U_+ + 2 U_-)\,\Delta - \frac{1}{8m}\,\Delta(U_+ + 2U_-)
- \frac{i}{2m}\,\vec{S}\cdot \Big(\vec{\nabla}(U_+ + 2 U_-) \times
\vec{\nabla}\,\Big)\nonumber\\ && + \frac{i}{2m}\,{\cal
  K}\,\vec{S}\cdot \vec{\nabla} + \frac{i}{4m}\,\vec{S}\cdot
\vec{\nabla}{\cal K},
\end{eqnarray}
reproduces the results, obtained in \cite{Ivanov2015a}.  The potentials
$\Phi^{(3)}_{\rm eff}(t,\vec{r},\vec{S}\,)$  and $\Phi^{(4)}_{\rm
  eff}(t,\vec{r},\vec{S}\,)$  are equal to
\begin{eqnarray}\label{eq:21}
\hspace{-0.3in}\Phi^{(3)}_{\rm eff}(t,\vec{r},\vec{S}\,) = -
\frac{i}{6m}\,(\vec{K}\cdot \vec{\cal B}\,)\,\vec{S}\cdot \vec{\nabla}
- \frac{i}{12 m}\,\vec{S}\cdot \vec{\nabla}(\vec{K}\cdot \vec{\cal
  B}\,) -
\frac{i}{4m}\,(\epsilon^{\hat{j}\hat{k}\hat{\ell}}\,K_{\hat{j}}\,{\cal
  M}_{\hat{0}\hat{k}\hat{\ell}})\,\vec{S}\cdot \vec{\nabla} -
\frac{i}{8 m}\,\vec{S}\cdot
\vec{\nabla}(\epsilon^{\hat{j}\hat{k}\hat{\ell}}\,K_{\hat{j}}\,{\cal
  M}_{\hat{0}\hat{k}\hat{\ell}})
\end{eqnarray}
and
\begin{eqnarray}\label{eq:22}
\Phi^{(4)}_{\rm eff}(t,\vec{r},\vec{S}\,) &=& - \frac{2}{3m}\,{\cal
  E}_{\hat{0}}\,\vec{K}\cdot \vec{\nabla} -
\frac{1}{3m}\,\vec{\nabla}\cdot ({\cal
  E}_{\hat{0}}\,\vec{K}\,) - \frac{1}{2m}\,({\cal
  M}_{\hat{\ell}\hat{j}\hat{0}} + {\cal M}_{\hat{0}\hat{j}\hat{\ell}}
) \,K^{\hat{\ell}}\frac{\partial}{\partial x_{\hat{j}}} -
\frac{1}{4m}\,\frac{\partial}{\partial x_{\hat{j}}}\Big(({\cal
  M}_{\hat{\ell}\hat{j}\hat{0}} + {\cal M}_{\hat{0}\hat{j}\hat{\ell}}
)\,K^{\hat{\ell}}\Big) \nonumber\\ && + \frac{2}{3m}\,i\,\vec{S}\cdot
     {\rm rot}({\cal E}_{\hat{0}}\,\vec{K}\,) -
     \frac{1}{2m}\,i\,\epsilon^{\hat{j}\hat{k}\hat{\ell}}\,S_{\hat{\ell}}\,
     \frac{\partial}{\partial x^{\hat{j}}}\Big(({\cal
       M}_{\hat{a}\hat{j}\hat{0}} + {\cal M}_{\hat{0}\hat{j}\hat{a}} )
     \,K^{\hat{a}}\Big),
\end{eqnarray}
and define new torsion--fermion (neutron) low--energy interactions. On
the whole the effective low--energy potential Eq.(\ref{eq:18})
possesses the following properties. First of all, we would like to
accentuate that to linear approximation in gravitational, chameleon
and torsion fields there are no new chameleon--fermion interactions in
comparison with those, calculated in \cite{Ivanov2015a}. To order
$1/m$ the potential $\Phi^{(3)}_{\rm eff}(t,\vec{r},\vec{S}\,)$
describes new torsion--fermion interactions of torsion axial--vector
$\vec{\cal B}$ and tensor ${\cal M}_{\hat{0}\hat{k}\hat{\ell}}$
degrees of freedom. Then, the effective low--energy potential
$\Phi^{(4)}_{\rm eff}(t,\vec{r},\vec{S}\,)$, containing new torsion
interactions with slow fermions to order $1/m$, is anti--hermitian. It
violates invariance under time reversal transformation (or
$T$--invariance) and under Charge--Parity transformation (or
$CP$--invariance). In section \ref{sec:Hermiticity} we
discuss in detail such a property of the effective low--energy
potential $\Phi^{(4)}_{\rm eff}(t,\vec{r},\vec{S}\,)$.

\section{Spacetime metric and anti--Hermiticity of the effective 
low--energy potential $\Phi^{(4)}_{\rm eff}(t,\vec{r},\vec{S}\,)$}
\label{sec:Hermiticity}

First of all we would like to note that the metric Eq.(\ref{eq:11}) is
not invariant under time reversal transformation $t \to - t$. As a
result, one can expect a possible violation of $T$--invariance
\cite{Hadley2011}. Since a violation of time reversal invariance is
yielded by the effective low--energy potential $\Phi^{(4)}_{\rm
  eff}(t,\vec{r},\vec{S}\,)$, below we analyse step by step the
appearance of such a potential.

The operator $\Phi^{(4)}_{\rm eff}(t,\vec{r},\vec{S}\,)$ one may
obtain from the following terms of the effective low--energy potential
Eq.(\ref{eq:17})
\begin{eqnarray}\label{eq:23} 
\delta \Phi_{\rm eff}(t, \vec{r},\vec{S}\,) &=& \frac{1}{2 m
  A}\,\eta^{\hat{j}\hat{k}}\,G_{\hat{j}}\,D^k_{\hat{k}}\,
\frac{\partial}{\partial x^k} + \frac{1}{4 m
}\,\eta^{\hat{j}\hat{k}}\,D^k_{\hat{k}}\, \frac{\partial}{\partial
  x^k}\Big(\frac{G_{\hat{j}}}{A}\Big) - \frac{1}{2 m
}\,i\,\epsilon^{\hat{j}\hat{k}\hat{\ell}}\,
S_{\hat{\ell}}\,D^k_{\hat{k}}\,\frac{\partial}{\partial
  x^k}\Big(\frac{G_{\hat{j}}}{A}\Big) \nonumber\\
\hspace{-0.3in}&+& \frac{1}{2 m
  A}\,\eta^{\hat{j}\hat{k}}\,G_{\hat{k}}\,D^j_{\hat{j}}
\frac{\partial}{\partial x^j} + \frac{1}{4 m
  A}\,\eta^{\hat{j}\hat{k}}\,D^j_{\hat{j}}\frac{\partial
  G_{\hat{k}}}{\partial x^j} + \frac{1}{2 m
  A}\,i\,\epsilon^{\hat{j}\hat{k}\hat{\ell}}\,
S_{\hat{\ell}}\,D^j_{\hat{j}}\frac{\partial G_{\hat{k}}}{\partial
  x^j}.
\end{eqnarray}
In the linear approximation for interacting gravitational, chameleon
and torsion fields the potential Eq.(\ref{eq:23}) reduces to the form
\begin{eqnarray}\label{eq:24} 
\delta \Phi_{\rm eff}(t, \vec{r},\vec{S}\,) = -
\frac{1}{m}\,\eta^{\hat{j}\hat{k}}\,G_{\hat{j}}\,
\frac{\partial}{\partial x^{\hat{k}}} - \frac{1}{2 m
}\,\eta^{\hat{j}\hat{k}}\, \frac{\partial G_{\hat{j}}}{\partial
  x^{\hat{k}}} + \frac{1}{m
}\,i\,\epsilon^{\hat{j}\hat{k}\hat{\ell}}\,
S_{\hat{\ell}}\,\frac{\partial G_{\hat{k}}}{\partial x^{\hat{j}}},
\end{eqnarray}
where we have replaced $A \to 1$, $D^k_{\hat{k}} \to -
\delta^k_{\hat{k}}$ and $D^j_{\hat{j}} \to - \delta^j_{\hat{j}}$ (see
Eq.(\ref{eq:14})). Setting $K^{\hat{j}} = 0$ we get  (see the Appendix)
\begin{eqnarray}\label{eq:25}
G_{\hat{j}} = - \frac{1}{2}\,\frac{\partial}{\partial x^{\hat{j}}}(U_+
+ U_-).
\end{eqnarray}
The contribution of the operator $G_{\hat{j}}$, given by
Eq.(\ref{eq:25}), is important for the derivation of the
gravitational--chameleon part of the effective low--energy potential
$\Phi^{(2)}_{\rm eff}(t, \vec{r},\vec{S}\,)$, which is Hermitian. This
confirms the correctness of the terms proportional to $G_{\hat{j}}$
and $G_{\hat{k}}$ in the effective low--energy potential $\Phi_{\rm
  eff}(t, \vec{r}, \vec{S}\,)$.

In the curved spacetime with torsion and metric Eq.(\ref{eq:11}) the
operator $G_{\hat{j}}$ acquires a certain contribution of the torsion
field (see the Appendix)
\begin{eqnarray}\label{eq:26}
G_{\hat{j}} = - \frac{1}{2}\,\frac{\partial}{\partial x^{\hat{j}}}(U_+
+ U_-) + \frac{2}{3}\,{\cal E}_{\hat{0}}\,K_{\hat{j}} + \frac{1}{2}\,
({\cal M}_{\hat{\ell}\hat{j}\hat{0}} + {\cal
  M}_{\hat{0}\hat{j}\hat{\ell}} )\,K^{\hat{\ell}}.
\end{eqnarray}
Another confirmation of the correctness of the calculation of the
contribution of torsion field to the operator $G_{\hat{j}}$ is a
cancellation of the part independent of $K^{\hat{j}}$. Indeed, a direct
calculation of the torsion $K^{\hat{j}}$--independent part in the operator
$G_{\hat{j}}$ gives (see the Appendix)
\begin{eqnarray}\label{eq:27}
G_{\hat{j}}\Big|^{\rm torsion}_{K^{\hat{j}} = 0} = \frac{1}{2}\,{{\cal
    T}^{\hat{\alpha}}}_{\hat{\alpha}\hat{j}} + {\cal
  K}_{\hat{j}\hat{0}\hat{0}} + {\cal
  K}_{\hat{j}\hat{\ell}\hat{k}}\,\eta^{\hat{\ell}\hat{k}} =
\frac{1}{2}\,{\cal T}_{\hat{0}\hat{0}\hat{j}} + \frac{1}{2}\,{\cal
  T}_{\hat{j}\hat{\ell}\hat{k}}\,\eta^{\hat{\ell}\hat{k}} + \frac{1}{2}\,{\cal
  K}_{\hat{j}\hat{0}\hat{0}} + \frac{1}{2}\,{\cal
  K}_{\hat{j}\hat{\ell}\hat{k}}\,\eta^{\hat{\ell}\hat{k}} = 0.
\end{eqnarray}
In the right--hand--side (r.h.s.) of Eq.(\ref{eq:27}) the first and
second terms are cancelled by the third and fourth ones,
respectively. This agrees well with results, obtained by Kostelecky
\cite{Kostelecky2004} (see also \cite{Ivanov2015a}). In turn, the contribution of the $K^{\hat{j}}$--dependent part (see the Appendix)
\begin{eqnarray}\label{eq:28}
G_{\hat{j}}\Big|^{\rm torsion}_{O(K^{\hat{j}})} &=&
\frac{1}{2}\,({\cal K}_{\hat{j}\hat{0}\hat{\ell}} + {\cal
  K}_{\hat{j}\hat{\ell}\hat{0}})\,K^{\hat{\ell}} -
\frac{1}{2}\,K_{\hat{j}}\,{\cal
  K}_{\hat{0}\hat{\ell}\hat{k}}\,\eta^{\hat{\ell}\hat{k}} =
\frac{1}{2}\,({\cal T}_{\hat{0}\hat{j}\hat{\ell}} + {\cal
  T}_{\hat{\ell}\hat{j}\hat{0}}) K^{\hat{\ell}} +
\frac{1}{2}\,K_{\hat{j}}\,{\cal
  T}_{\hat{\ell}\hat{k}\hat{0}}\,\eta^{\hat{\ell}\hat{k}} =
\nonumber\\ &=& \frac{2}{3}\,{\cal E}_{\hat{0}}\,K_{\hat{j}} +
\frac{1}{2}\,({\cal M}_{\hat{\ell}\hat{j}\hat{0}} + {\cal
       M}_{\hat{0}\hat{j}\hat{\ell}} )\,K^{\hat{\ell}} \neq 0,
\end{eqnarray}
which is fully correlated with the $K^{\hat{j}}$--independent part
through the affine connection (see the Appendix), does not
vanish. Hence, taking the operator $G_{\hat{j}}$, given by
Eq.(\ref{eq:26}), we obtain the effective low--energy potential
$\Phi^{(4)}_{\rm eff}(t,\vec{r},\vec{S}\,)$, which is anti--Hermitian
only due to the contribution of torsion and contains both
torsion--non--spin--matter and torsion--spin--matter
interactions. Thus, we have proved that the appearance of the
anti--Hermitian potential $\Phi^{(4)}_{\rm eff}(t, \vec{r},
\vec{S}\,)$ is not a mistake of the calculation but an objective
reality, caused by the presence of the torsion scalar ${\cal
  E}_{\hat{0}}$ and tensor ${\cal M}_{\hat{\ell}\hat{j}\hat{0}}$ and
${\cal M}_{\hat{0}\hat{j}\hat{\ell}}$ degrees of freedom in rotating
coordinate systems.

\section{Heisenberg's equation for spin operator of slow neutrons in 
Einstein--Cartan gravity with torsion, chameleon and magnetic fields}
\label{sec:equation}

A time evolution of the neutron spin operator $\vec{S}$ is described
by Heisenberg's equation of motion \cite{LL2007}
\begin{eqnarray}\label{eq:29}
\frac{d\vec{S}}{dt} = \frac{\partial \vec{S}}{\partial t}
+ i\,[{\rm H}, \vec{S}\,].
\end{eqnarray}
Since the spin operator $\vec{S}$ does not depend explicitly on time,
the partial derivative in Eq.(\ref{eq:29}) is equal to zero. This
yields
\begin{eqnarray}\label{eq:30}
\frac{d\vec{S}}{dt} = i\,[{\rm H}, \vec{S}\,].
\end{eqnarray}
Since the operator of the kinetic energy of slow neutrons commutes with a neutron spin, we arrive at the equation
\begin{eqnarray}\label{eq:31}
\frac{dS^{\hat{a}}}{dt} =
\epsilon^{\hat{a}\hat{b}\hat{c}}\,\Omega_{\rm m \hat{b}}S_{\hat{c}} +
i\,[ \tilde{\Phi}_{\rm eff}(t, \vec{r}, \vec{S}\,), S^{\hat{a}}],
\end{eqnarray}
where $\Omega_{m \hat{b}} = - \kappa_n\mu_N B_{\hat{b}}$ with
$B_{\hat{b}} = (- \vec{B}\,)_{\hat{b}}$ (or $\vec{\Omega}_m = -
\kappa_n\mu_N\,\vec{B}$), is the standard angular velocity of the
neutron spin precession in the magnetic field $\vec{B}$
\cite{Sears1989}.  For the calculation of the term
$\epsilon^{\hat{a}\hat{b}\hat{c}}\,\Omega_{ m \hat{b}}S_{\hat{c}} =
(\vec{\Omega}_m\times \vec{S}\,)^{\hat{a}}$, where $\Omega_{m \hat{b}}
= (- \vec{\Omega}_m)_{\hat{b}}$ and $S_{\hat{c}} = (-
\vec{S}\,)_{\hat{c}}$, we have used the commutation relation
$[S^{\hat{b}}, S^{\hat{a}}] =
i\,\epsilon^{\hat{a}\hat{b}\hat{c}}\,S_{\hat{c}}$. The contribution of
the commutator $[ \Phi_{\rm eff}(t, \vec{r}, \vec{S}\,), S^{\hat{a}}]$
can be written in the following standard form
\begin{eqnarray}\label{eq:32}
i\,[\tilde{\Phi}_{\rm eff}(t, \vec{r}, \vec{S}\,), S^{\hat{a}}] =
\epsilon^{\hat{a}\hat{b}\hat{c}}\Omega_{\hat{b}} S_{\hat{c}} =
(\vec{\Omega}\times \vec{S}\,)^{\hat{a}},
\end{eqnarray}
where $\vec{\Omega}$ is the angular velocity operator of the neutron
spin precession, determined by
\begin{eqnarray}\label{eq:33}
\vec{\Omega} = \vec{\Omega}_{\rm r} + \vec{\Omega}_{\rm gr-ch} +
\vec{\Omega}_{\rm t} + \vec{\Omega}_{\rm \bar{h}},
\end{eqnarray}
where the indices ${\rm r}$, ${\rm gr-ch}$, ${\rm t}$ and $\bar{h}$
mean ``rotation'',''gravitation--chameleon'', ``torsion'' and
``anti--Hermitian'', respectively.  The angular velocity operators in
the r.h.s. of Eq.(\ref{eq:33}) are equal to
\begin{eqnarray}\label{eq:34}
\Omega^{\hat{b}}_{\rm r} &=& -
\frac{1}{2}\,\epsilon^{\hat{b}\hat{k}\hat{\ell}}\frac{\partial
}{\partial x^{\hat{k}}} K_{\hat{\ell}},
\nonumber\\ \Omega^{\hat{b}}_{\rm gr-ch} &=& -
\frac{i}{2m}\,\epsilon^{\hat{b}\hat{j}\hat{k}}\,\frac{\partial}{\partial
  x^{\hat{j}}}(U_+ + 2 U_-)\,\frac{\partial}{\partial
  x^{\hat{k}}},\nonumber\\ \Omega^{\hat{b}}_{\rm t} &=& -
\frac{1}{2}\,{\cal B}^{\hat{b}} - \frac{1}{3}\,{\cal K}\,K^{\hat{b}} +
\frac{1}{2}\,\epsilon^{\hat{b}\hat{j}\hat{k}}\,K_{\hat{j}}{\cal
  M}_{\hat{0}\hat{0}\hat{k}} -
\frac{1}{2}\,\epsilon^{\hat{b}\hat{k}\hat{\ell}}\,{\cal
  M}_{\hat{k}\hat{\ell}\hat{j}}\,K^{\hat{j}} - \frac{i}{2m}\,{\cal
  K}\,\frac{\partial}{\partial x_{\hat{b}}} -
\frac{i}{4m}\,\frac{\partial}{\partial x_{\hat{b}}}{\cal K}
\nonumber\\ && + \frac{i}{6m}\,(\vec{K}\cdot \vec{\cal
  B}\,)\,\frac{\partial}{\partial x_{\hat{b}}} + \frac{i}{12
  m}\,\frac{\partial}{\partial x_{\hat{b}}}(\vec{K}\cdot \vec{\cal
  B}\,) +
\frac{i}{4m}\,(\epsilon^{\hat{j}\hat{k}\hat{\ell}}\,K_{\hat{j}}\,{\cal
  M}_{\hat{0}\hat{k}\hat{\ell}})\,\frac{\partial}{\partial
  x_{\hat{b}}} + \frac{i}{8 m}\,\frac{\partial}{\partial
  x_{\hat{b}}}(\epsilon^{\hat{j}\hat{k}\hat{\ell}}\,K_{\hat{j}}\,{\cal
  M}_{\hat{0}\hat{k}\hat{\ell}}),\nonumber\\ \Omega^{\hat{b}}_{\rm
  \bar{h}} &=& -
\frac{2}{3m}\,i\,\epsilon^{\hat{b}\hat{k}\hat{\ell}}\frac{\partial
}{\partial x^{\hat{k}}}({\cal E}_{\hat{0}}K_{\hat{\ell}}) +
\frac{1}{2m}\,i\,\epsilon^{\hat{b}\hat{k}\hat{\ell}}\,
\frac{\partial}{\partial x^{\hat{\ell}}}\Big(({\cal
  M}_{\hat{j}\hat{k}\hat{0}} + {\cal M}_{\hat{0}\hat{k}\hat{j}} )
\,K^{\hat{j}}\Big).
\end{eqnarray}
For the coordinate system, rotating with an angular velocity
$\vec{\omega}$, where  $K^{\hat{j}} = -
\epsilon^{\hat{j}\hat{k}\hat{\ell}}\omega_{\hat{k}}x_{\hat{\ell}} = -
(\vec{\omega}\times \vec{r}\,)^{\hat{j}}$, we get
\begin{eqnarray}\label{eq:35}
\Omega^{\hat{b}}_{\rm r} &=& - \omega^{\hat{b}},
\nonumber\\ \Omega^{\hat{b}}_{\rm gr-ch} &=& -
i\,\frac{3}{2m}\,(\vec{\nabla}U_{\rm E}\times
\vec{\nabla}\,)^{\hat{b}} + i\,\frac{1}{2m}\,\frac{\beta}{M_{\rm
    Pl}}\,(\vec{\nabla}\phi \times
\vec{\nabla}\,)^{\hat{b}},\nonumber\\ \Omega^{\hat{b}}_{\rm t} &=& -
\frac{1}{2}\,{\cal B}^{\hat{b}} + \frac{1}{3}\,(\vec{\omega}\times
\vec{r}\,{\cal K})^{\hat{b}} -
\frac{1}{2}\,\Big(x^{\hat{b}}(\omega^{\hat{k}}{\cal
  M}_{\hat{0}\hat{0}\hat{k}}) - \omega^{\hat{b}}(x^{\hat{k}}{\cal
  M}_{\hat{0}\hat{0}\hat{k}})\Big) +
\frac{1}{2}\,\epsilon^{\hat{b}\hat{k}\hat{\ell}}\,{\cal
  M}_{\hat{k}\hat{\ell}\hat{j}}\,\epsilon^{\hat{j}\hat{p}\hat{q}}\,\omega_{\hat{p}}\,
x_{\hat{q}}\nonumber\\ &&- \frac{i}{2m}\,{\cal
  K}\,\frac{\partial}{\partial x_{\hat{b}}} -
\frac{i}{4m}\,\frac{\partial}{\partial x_{\hat{b}}}{\cal K} -
\frac{i}{6m}\,\Big(\vec{\omega}\cdot (\vec{r}\times \vec{\cal
  B}\,)\Big)\,\frac{\partial}{\partial x_{\hat{b}}} - \frac{i}{12
  m}\,\frac{\partial}{\partial x_{\hat{b}}}\Big(\vec{\omega}\cdot
(\vec{r}\times \vec{\cal B}\,)\Big)\nonumber\\ && +
\frac{i}{4m}\,\Big((\omega^{\hat{k}}x^{\hat{\ell}} -
\omega^{\hat{\ell}}x^{\hat{k}})\,{\cal
  M}_{\hat{0}\hat{k}\hat{\ell}}\Big)\,\frac{\partial}{\partial
  x_{\hat{b}}} + \frac{i}{8 m}\,\frac{\partial}{\partial
  x_{\hat{b}}}\Big((\omega^{\hat{k}}x^{\hat{\ell}} -
\omega^{\hat{\ell}}x^{\hat{k}})\,{\cal
  M}_{\hat{0}\hat{k}\hat{\ell}}\Big),\nonumber\\ \Omega^{\hat{b}}_{\rm
  \bar{h}} &=& - \frac{2}{3m}\,i\,\Big(\vec{\nabla} {\cal E}_{\hat{0}}
\times (\vec{\omega}\times \vec{r}\,) + 2 {\cal
  E}_{\hat{0}}\,\vec{\omega}\,\Big)^{\hat{b}} +
\frac{1}{2m}\,i\,\epsilon^{\hat{b}\hat{k}\hat{\ell}}\,
\frac{\partial}{\partial x^{\hat{\ell}}}\Big(({\cal
  M}_{\hat{j}\hat{k}\hat{0}} + {\cal M}_{\hat{0}\hat{k}\hat{j}} )
\,\epsilon^{\hat{j}\hat{p}\hat{q}}\omega_{\hat{p}}x_{\hat{q}}\Big).
\end{eqnarray}
The experimental analysis of the spin--rotation couplings, calculated
above, can be, for example, carried out by neutron interferometer
\cite{Rauch2000}--\cite{Demirel2015} with rotating interferometers
\cite{Atwood1984,Mashhoon1988,Demirel2015}.

\section{Properties of the Dirac Hamilton operator in
    curved spacetimes}
\label{sec:obukhov}

The anti--Hermitian interactions appears also in the Hamilton operator
of a relativistic Dirac fermion with mass $m$. Indeed, as has been
shown in \cite{Ivanov2015b} the Hamilton operator of a relativistic
Dirac fermion is equal to ${\rm H}' = {\rm H}_0 + \delta {\rm H}'$,
where ${\rm H}_0 = \gamma^{\hat{0}} m -
i\,\gamma^{\hat{0}}\vec{\gamma}\cdot \vec{\nabla}$ is the Hamilton
operator of a free Dirac fermion with mass $m$ and $\delta {\rm H}'$
is the interaction Hamilton operator equal to \cite{Ivanov2015b}
\begin{eqnarray}\label{eq:36}
\delta {\rm H}' = (A - 1) \gamma^{\hat{0}}m + B +
C^{\hat{\ell}}\Sigma_{\hat{\ell}} + (D^j_{\hat{j}} +
\delta^j_{\hat{j}})\,i\gamma^{\hat{0}}\gamma^{\hat{j}}\frac{\partial}{\partial
  x^j} +
F_{\hat{j}}i\gamma^{\hat{0}}\gamma^{\hat{j}}\frac{\partial}{\partial
  t} + G_{\hat{j}}i\gamma^{\hat{0}}\gamma^{\hat{j}} + K\,\gamma^5 +
L^{\hat{j}}\,i\,\frac{\partial}{\partial x^{\hat{j}}}.
\end{eqnarray}
For the operators $A$, $B$, $C^{\hat{\ell}}$, $D^j{\hat{j}}$,
$G_{\hat{j}}$, $K$ and $L^{\hat{j}}$, given by Eq.(\ref{eq:14}), the Hamilton
operator can be represented in the form $\delta {\rm H}' = \delta {\rm
  H}'_{\rm h} + \delta {\rm H}'_{\rm \bar{h}}$, where $\delta {\rm
  H}'_{\rm h}$ and $\delta {\rm H}'_{\rm \bar{h}}$ are the Hermitian and
anti--Hermitian parts of the Hamilton operator Eq.(\ref{eq:36}),
respectively, equal to
\begin{eqnarray}\label{eq:37}
\delta {\rm H}'_{\rm h} &=& U_+ \gamma^{\hat{0}}m - (U_+ +
U_-)\,i\,\gamma^{\hat{0}}\vec{\gamma}\cdot \vec{\nabla} -
\frac{1}{2}\,i\,\gamma^{\hat{0}}\,\vec{\nabla}(U_+ +
U_-)\nonumber\\ &-& \frac{1}{2}\,i\,{\rm div}\,\vec{K} -
i\,\vec{K}\cdot \vec{\nabla} + \frac{1}{4}\,\vec{\Sigma} \cdot {\rm
  rot}\,\vec{K} \nonumber\\ &-& \frac{1}{4}\, \vec{\Sigma} \cdot
\vec{\cal B} - \frac{1}{6}\,{\cal K}\,\vec{\Sigma} \cdot \vec{K} +
\frac{1}{4}\,\epsilon^{\hat{\ell} \hat{j}
  \hat{k}}\,\Sigma_{\hat{\ell}}\,K_{\hat{j}}\,{\cal
  M}_{\hat{0}\hat{0}\hat{k}} + \frac{1}{4}\,\epsilon^{\hat{\ell}
  \hat{j} \hat{k}}\,\Sigma_{\hat{\ell}}\,{\cal
  M}_{\hat{j}\hat{k}\hat{a}}\,K^{\hat{a}}\nonumber\\ &-&
\frac{1}{4}\,{\cal K}\,\gamma^5 + \frac{1}{12}\,\vec{K}\cdot \vec{\cal
  B}\,\gamma^5 +
\frac{1}{8}\,\epsilon^{\hat{j}\hat{k}\hat{\ell}}\,K_{\hat{j}}\,{\cal
  M}_{\hat{0}\hat{k}\hat{\ell}}\,\gamma^5
\end{eqnarray}
and 
\begin{eqnarray}\label{eq:38}
\delta {\rm H}'_{\rm \bar{h}} = - i\,\frac{2}{3}\,{\cal
  E}_{\hat{0}}\,\vec{K}\cdot \gamma^{\hat{0}}\vec{\gamma} +
i\,\frac{1}{2}\,({\cal M}_{\hat{\ell}\hat{j}\hat{0}} + {\cal
  M}_{\hat{0}\hat{j}\hat{\ell}}
)\,K^{\hat{\ell}}\gamma^{\hat{0}}\gamma^{\hat{j}}.
\end{eqnarray}
One may assume that in the rotating Universe and galaxies
\cite{Gamow1946} (see also \cite{Obukhov1997}) the torsion--fermion
interaction Eq.(\ref{eq:38}) might be an origin of i) violation of
$CP$ and $T$ invariance in the Universe and ii) of baryon asymmetry
\cite{PDG2014}.

\subsection{Standard non--unitary transformation of
    Dirac fermion wave functions and anti--Hermitian torsion--fermion
    interactions}

It is well--known that the Hamilton operator of the
  Dirac fermions with mass $m$, moving in the curved spacetime with a
  metric $g_{\mu\nu}$, is not Hermitian. In order to get a Hermitian
  Hamilton operator one has to perform the standard non--unitary
  transformation of the wave function of the Dirac fermions $\psi \to
  (\sqrt{- g}\,e^{\hat{0}}_0)^{1/2} \psi'$, where $g$ is a determinant
  of the metric tensor $g_{\mu\nu}(x)$ and $e^{\hat{0}}_0(x)$ is a
  vierbein field \cite{Fischbach1981,Obukhov2001},
  \cite{Obukhov2009,Obukhov2011,Obukhov2013}, \cite{Obukhov2014},
  \cite{Silenko2013,Silenko2015} and \cite{Gorbatenko2010} (see also
  \cite{Ivanov2015a,Ivanov2015b,Ivanov2014,Ivanov2015c}).  As has been
  shown in \cite{Ivanov2015b}, the Hamilton operator ${\rm H}' =
  \gamma^{\hat{0}} m - i\,\gamma^{\hat{0}}\vec{\gamma}\cdot
  \vec{\nabla} + \delta {\rm H}'_{\rm h} + \delta {\rm H}'_{\rm \bar{h}}$
  has been already obtained by means of the standard non--unitary
  transformation $\psi \to (\sqrt{-
    \tilde{g}}\,\tilde{e}^{\hat{0}}_0)^{1/2} \psi'$, where $\tilde{g}$
  is a determinant of the Jordan--frame metric tensor
  $\tilde{g}_{\mu\nu}(x)$ and $\tilde{e}^{\hat{0}}_0(x)$ is a vierbein
  field in the Jordan frame (see Eq.(\ref{eq:13}) of
  Ref.\cite{Ivanov2015b}). The appearance of the anti--Hermitian term
  $\delta {\rm H}'_{\rm \bar{h}}$ in the Hamilton operator ${\rm H}' =
  \gamma^{\hat{0}} m - i\,\gamma^{\hat{0}}\vec{\gamma}\cdot
  \vec{\nabla} + \delta {\rm H}'_{\rm h} + \delta {\rm H}'_{\rm \bar{h}}$
  is fully related to the spacetime metric Eq.(\ref{eq:11}) as a
  functional of the vector $\vec{K}$, caused by rotations, or more
  generally to the spacetime metric Eq.(\ref{eq:3}) (see also
  Eq.(\ref{eq:20}) of Ref.\cite{Ivanov2015b}), proposed by Obukhov,
  Silenko, and Teryaev \cite{Obukhov2009,Obukhov2011,Obukhov2014}.

\subsection{Non--unitary transformations of Dirac
    fermion wave functions and removal of anti--Hermitian
    torsion--fermion interactions}

Now we would like to show that the anti--Hermitian (non--Hermitian)
Hamilton operator $\delta {\rm H}'_{\rm \bar{h}}$, given by
Eq.(\ref{eq:38}), cannot be removed by a non--unitary (non--Hermitian)
transformation of the Dirac fermion (neutron) wave function. After the
standard non--unitary transformation of the Dirac fermion wave
function $\psi \to (\sqrt{- \tilde{g}}\,\tilde{e}^{\hat{0}}_0)^{1/2}
\psi'$ (see Eq.(\ref{eq:13}) of Ref.\cite{Ivanov2015b}) we arrive at
the following Dirac fermion action
\begin{eqnarray}\label{eq:39}
S_{\psi} = \int
dt d^3x\,\psi'^{\dagger}(t,\vec{r}\,)\Big(i\frac{\partial }{\partial t}
- {\rm H}'\Big)\psi'(t,\vec{r}\,) = \int
dt d^3x\,\psi'^{\dagger}(t,\vec{r}\,)\Big(i\frac{\partial }{\partial t}
- {\rm H}_0 - \delta {\rm H}'_{\rm h} - \delta {\rm
  H}'_{\rm \bar{h}}\Big)\psi'(t,\vec{r}\,),
\end{eqnarray}
where ${\rm H}_0 = \gamma^{\hat{0}} m -
i\,\gamma^{\hat{0}}\vec{\gamma}\cdot \vec{\nabla}$.  In order to
analyse a possibility to remove the term $\delta {\rm H}'_{\rm \bar{h}}$
we make a non--unitary (non--Hermitian) transformation
\begin{eqnarray}\label{eq:40}
\psi'(t,\vec{r}\,) = \zeta\,\psi''(t,\vec{r}\,),
\end{eqnarray}
where $\zeta = 1 + Q = 1 + \eta_{\hat{j}\hat{k}}
O^{\hat{j}}\gamma^{\hat{k}}$ and $O^{\hat{j}}$ is a Hermitian
non--differential operator, i.e. $Q^{\hat{j} \dagger} =
Q^{\hat{j}}$. The operator $\zeta$ is a non--unitary (non--Hermitian)
operator $\zeta^{\dagger} = 1 + Q^{\dagger} = 1 -
\eta_{\hat{j}\hat{k}} O^{\hat{j}}\gamma^{\hat{k}} \neq \zeta$. Then,
for the derivation of the Dirac Hamilton operator ${\rm H}''$ we use
the following relations $\zeta^{\dagger}\delta {\rm H}'_{\rm h} =
\delta {\rm H}'_{\rm h} $ and of $\zeta^{\dagger} \delta {\rm H}'_{\rm
  \bar{h}} = \delta {\rm H}'_{\rm \bar{h}} $ and $\delta {\rm H}'_{\rm
  h} \zeta = \delta {\rm H}'_{\rm h}$ and $\delta {\rm H}'_{\rm
  \bar{h}} \zeta = \delta {\rm H}'_{\rm \bar{h}}$. Plugging Eq.(\ref{eq:40}) into Eq.(\ref{eq:39}) we
transcribe the action $S_{\psi}$ into the form
\begin{eqnarray}\label{eq:41}
S_{\psi} = \int
dt d^3x\,\psi''^{\dagger}(t,\vec{r}\,)\Big(i\frac{\partial }{\partial t}
- {\rm H}''\Big)\psi''(t,\vec{r}\,),
\end{eqnarray}
where the Hamilton operator ${\rm H}''$ is equal to
\begin{eqnarray}\label{eq:42}
{\rm H}'' = {\rm H}_0 + \delta {\rm H}'_{\rm h} + \delta {\rm
  H}'_{\rm \bar{h}} + [{\rm H}_0, Q] - i\frac{\partial Q}{\partial t}.
\end{eqnarray}
Plugging $Q = \eta_{\hat{j}\hat{k}} O^{\hat{j}}\gamma^{\hat{k}}$ into
Eq.(\ref{eq:42}) and calculating the commutator $[{\rm H}_0, Q]$ we
get
\begin{eqnarray}\label{eq:43}
{\rm H}'' = {\rm H}_0 + \delta {\rm H}'_{\rm h} + \delta {\rm
  H}'_{\rm \bar{h}} + 2 m \eta_{\hat{j}\hat{k}}
Q^{\hat{j}}\gamma^{\hat{0}}\gamma^{\hat{k}} -
i\,\gamma^{\hat{0}}\frac{\partial Q^{\hat{j}}}{\partial x^{\hat{j}}} -
2 i\,\gamma^{\hat{0}} Q^{\hat{j}}\frac{\partial }{\partial
  x^{\hat{j}}} +
\varepsilon^{\hat{j}\hat{\ell}\hat{k}}\gamma^{\hat{0}}\Sigma_{\hat{j}}\frac{\partial
  Q_{\hat{k}}}{\partial x^{\hat{\ell}}} - i\,\frac{\partial
  Q_{\hat{j}}}{\partial t}\,\gamma^{\hat{j}}.
\end{eqnarray}
It is obvious that for the Hermitian operator $Q^{\hat{j}}$,
i.e. $Q^{\hat{j}\dagger} = Q^{\hat{j}}$ corresponding to a
non--Hermitian transformation with the operator $\zeta^{\dagger} \neq
\zeta$, the term $2 m \eta_{\hat{j}\hat{k}}
Q^{\hat{j}}\gamma^{\hat{0}}\gamma^{\hat{k}}$ cannot cancel the
contribution of the anti--Hermitian operator $\delta {\rm H}'_{\rm
  \bar{h}}$. The use of the anti--Hermitian operator $Q^{\hat{j}} \to
i Q^{\hat{j}}$, corresponding to a Hermitian (unitary) transformation
with an operator $\zeta = 1 + i\,Q = 1 + i\,\eta_{\hat{j}\hat{k}}
O^{\hat{j}}\gamma^{\hat{k}}$ such as $\zeta^{\dagger} = \zeta$, allows
to shift the Hamilton operator $\delta {\rm H}'_{\rm \bar{h}}$, which
is the {\it odd} operator according to the Foldy--Wouthuysen
classification \cite{Foldy1950}, to the region
of interactions of order $O(1/m)$. In detail such a unitary
transformation or the Foldy--Wouthuysen transformation for the
derivation of the effective low--energy potential Eq.(\ref{eq:17}) has
been performed in \cite{Ivanov2015b}. The linearised version of this
effective low--energy potential is given by Eq.(\ref{eq:18}).

\subsection{Non--Hermiticity of Dirac Hamilton
    operator, $\eta$--representation and anti--Hermitian
    torsion--fermion interactions}

An alternative transition from a non--Hermitian Hamilton operator of
the Dirac massive fermions, moving in the curved spacetime with an
arbitrary metric tensor $g_{\mu\nu}(x)$, to a Hermitian form can be
performed by using the $\eta$--representation of the Dirac fermion
wave functions \cite{Gorbatenko2010}. In the $\eta$--representation
the Dirac Hamilton operator becomes Hermitian without the standard
non--unitary transformation of the Dirac fermion wave function $\psi
\to (\sqrt{- \tilde{g}}\,\tilde{e}^0_{\hat{0}})^{1/2}\psi'$ and the
dynamics of the Dirac fermions is described by the pseudo--Hermitian
quantum mechanics \cite{Gorbatenko2010} (see also
\cite{Parker1980,Bender2007,Arminjon2011}). Since in our analysis of
the torsion--fermion interactions within the Einstein--Cartan gravity
with the chameleon field we use the standard non--unitary
transformation of the Dirac fermion wave function $\psi \to (\sqrt{-
  \tilde{g}}\,\tilde{e}^0_{\hat{0}})^{1/2}\psi'$, the dynamics of the
Dirac fermions and the contributions of the anti--Hermitian
torsion--fermion interactions, violating $CP$ and $T$ invariance, to
the observables can be described within the formalism of the standard
relativistic and non--relativistic quantum mechanics
\cite{Bjorken1966,Davydov1965}.

\subsection{Conformal invariance of anti--Hermitian
    torsion--fermion interactions}

As has been shown by Silenko \cite{Silenko2013}, the Dirac and
Foldy--Wouthuysen Hamilton operators for massless fermions in the
curved spacetimes with arbitrary metric $\tilde{g}_{\mu\nu}$ are
invariant under conformal transformation $\tilde{g}_{\mu\nu} \to O^2
\check{g}_{\mu\nu}$ if the wave function of massless fermions is
subjected to the non--unitary transformation $\psi' \to O^{3/2}
\check{\psi}$. In \cite{Silenko2015} the results, obtained in
\cite{Silenko2013}, have been extended to massive fermions coupled to
gravitational field and torsion in the Einstein and Einstein--Cartan
gravity, respectively, with the requirement that the fermion mass
transforms under the conformal transformation $\tilde{g}_{\mu\nu} \to
O^2 \check{g}_{\mu\nu}$ as follows $m \to O^{-1}\check{m}$. This
agrees well with the dimensional analysis of general relativity,
carried out by Dicke \cite{Dicke1962} for the reduction of the
Brans--Dicke gravitational theory \cite{Brans1961} to the Einstein
gravity, coupled to an effective scalar field. According to Silenko
\cite{Silenko2015}, the vector $K^j$ and the contorsion tensor
$\tilde{\cal K}_{\alpha\mu\nu} = - \frac{1}{2}( \tilde{\cal
  T}_{\alpha\mu\nu} - \tilde{\cal T}_{\mu\alpha\nu} - \tilde{\cal
  T}_{\nu\alpha\mu})$ are not changed by the conformal transformation
$\tilde{g}_{\mu\nu} \to O^2 \check{g}_{\mu\nu}$, i.e. $K^j \to
\check{K}^j = K^j$ and $\tilde{\cal K}_{\alpha\mu\nu} \to \check{\cal
  K}_{\alpha\mu\nu}$. Since the Hamilton operator $\delta {\rm
  H}_{\hat{h}}$ is expressed in terms of the components of the
contorsion tensor ${\cal K}_{\hat{\alpha}\hat{\mu}\hat{\nu}}$ and the
vector $K^{\hat{j}}$
\begin{eqnarray}\label{eq:44}
 \delta {\rm H}'_{\rm \bar{h}} &=& G_{\hat{j}}\Big|^{\rm
   torsion}_{O(K^{\hat{j}})} \,i\,\gamma^{\hat{0}}\gamma^{\hat{j}} =
 \frac{1}{2}\,({\cal K}_{\hat{j}\hat{0}\hat{\ell}} + {\cal
   K}_{\hat{j}\hat{\ell}\hat{0}})\,K^{\hat{\ell}}\,i\,\gamma^{\hat{0}}\gamma^{\hat{j}}
 - \frac{1}{2}\,K_{\hat{j}}\,{\cal
   K}_{\hat{0}\hat{\ell}\hat{k}}\,\eta^{\hat{\ell}\hat{k}}\,i\,
 \gamma^{\hat{0}}\gamma^{\hat{j}} =\nonumber\\ &=& -
 i\,\frac{2}{3}\,{\cal E}_{\hat{0}}\,\vec{K}\cdot
 \gamma^{\hat{0}}\vec{\gamma} + i\,\frac{1}{2}\,({\cal
   M}_{\hat{\ell}\hat{j}\hat{0}} + {\cal M}_{\hat{0}\hat{j}\hat{\ell}}
 )\,K^{\hat{\ell}}\gamma^{\hat{0}}\gamma^{\hat{j}},
\end{eqnarray}
it is invariant under the conformal transformation $\tilde{g}_{\mu\nu}
\to O^2 \check{g}_{\mu\nu}$, i.e. $\delta {\rm H}'_{\rm \bar{h}} \to \delta
\check{\rm H}'_{\rm \bar{h}} $, where $G_{\hat{j}}\Big|^{\rm
  torsion}_{O(K^{\hat{j}})}$, given by (see Eq.(\ref{eq:28}))
\begin{eqnarray}\label{eq:45}
G_{\hat{j}}\Big|^{\rm torsion}_{O(K^{\hat{j}})} = \frac{1}{2}\,({\cal
  K}_{\hat{j}\hat{0}\hat{\ell}} + {\cal
  K}_{\hat{j}\hat{\ell}\hat{0}})\,K^{\hat{\ell}} -
\frac{1}{2}\,K_{\hat{j}}\,{\cal
  K}_{\hat{0}\hat{\ell}\hat{k}}\,\eta^{\hat{\ell}\hat{k}},
\end{eqnarray}
is conformal invariant $G_{\hat{j}}\Big|^{\rm
  torsion}_{O(K^{\hat{j}})} \to \check{G}_{\hat{j}}\Big|^{\rm
  torsion}_{O(K^{\hat{j}})}$.  Under the conformal transformation
$\tilde{g}_{\mu\nu} \to O^2 \check{g}_{\mu\nu}$ the vierbein fields
transform as follows $\tilde{e}^{\hat{\alpha}}_{\mu} \to O
\check{e}^{\hat{\alpha}}_{\mu}$ and $\tilde{e}^{\mu}_{\hat{\alpha}}
\to O^{-1} \check{e}^{\mu}_{\hat{\alpha}}$. Since the fermion mass
transforms as $m \to O^{-1}\check{m}$, the operators $m A = - m
\tilde{e}^{\hat{0}}_0$ and $D^{j}_{\hat{j}} = - \tilde{e}^{\hat{0}}_0
\tilde{e}^j_{\hat{j}}$ (see Eq.(\ref{eq:7})) are invariant under the
conformal transformation, i.e. $m A \to \check{m} \check{A}$ and
$D^{j}_{\hat{j}} \to \check{D}^{j}_{\hat{j}}$. In order to show that
the effective low--energy potential $\Phi^{(4)}_{\rm eff}(t, \vec{r},
\vec{S}\,)$ is conformal invariant we transcribe the effective
low--energy potential $\delta \Phi_{\rm eff}(t, \vec{r},\vec{S}\,)$,
given by Eq.(\ref{eq:23}), into the form
\begin{eqnarray}\label{eq:46} 
\delta \Phi^{\rm torsion}_{\rm eff}(t, \vec{r},\vec{S}\,) &=& \frac{1}{2 m
  A}\,\eta^{\hat{j}\hat{k}}\,G_{\hat{j}}\Big|^{\rm
  torsion}_{O(K^{\hat{j}})}\,D^k_{\hat{k}}\, \frac{\partial}{\partial
  x^k} + \eta^{\hat{j}\hat{k}}\,D^k_{\hat{k}}\,
\frac{\partial}{\partial x^k}\Big(\frac{1}{4 m
  A}\,G_{\hat{j}}\Big|^{\rm torsion}_{O(K^{\hat{j}})}\Big) -
i\,\epsilon^{\hat{j}\hat{k}\hat{\ell}}\,
S_{\hat{\ell}}\,D^k_{\hat{k}}\,\frac{\partial}{\partial
  x^k}\Big(\frac{1}{2 m A}\,G_{\hat{j}}\Big|^{\rm
  torsion}_{O(K^{\hat{j}})}\Big) \nonumber\\
\hspace{-0.3in}&+& \frac{1}{2 m
  A}\,\eta^{\hat{j}\hat{k}}\,G_{\hat{k}}\Big|^{\rm
  torsion}_{O(K^{\hat{j}})}\,D^j_{\hat{j}} \frac{\partial}{\partial
  x^j} + \frac{1}{4 m
  A}\,\eta^{\hat{j}\hat{k}}\,D^j_{\hat{j}}\frac{\partial }{\partial
  x^j}G_{\hat{k}}\Big|^{\rm torsion}_{O(K^{\hat{j}})} + \frac{1}{2 m
  A}\,i\,\epsilon^{\hat{j}\hat{k}\hat{\ell}}\,
S_{\hat{\ell}}\,D^j_{\hat{j}}\frac{\partial }{\partial
  x^j}G_{\hat{k}}\Big|^{\rm torsion}_{O(K^{\hat{j}})},
\end{eqnarray}
Because of the conformal invariance of $m A \to \check{m} \check{A}$,
$D^j_{\hat{j}} \to \check{D}^j_{\hat{j}}$ and $G_{\hat{j}}\Big|^{\rm
  torsion}_{O(K^{\hat{j}})} \to \check{G}_{\hat{j}}\Big|^{\rm
  torsion}_{O(K^{\hat{j}})}$ the effective low--energy potential
Eq.(\ref{eq:46}) is conformal invariant, i.e. $\delta \Phi^{\rm
  torsion}_{\rm eff}(t, \vec{r},\vec{S}\,) \to \delta
\check{\Phi}^{\rm torsion}_{\rm eff}(t, \vec{r},\vec{S}\,)$. This
proves the conformal invariance of the effective low--energy potential
$\Phi^{(4)}_{\rm eff}(t, \vec{r},\vec{S}\,)$, i.e. $\Phi^{(4)}_{\rm
  eff}(t, \vec{r},\vec{S}\,) \to \check{\Phi}^{(4)}_{\rm eff}(t,
\vec{r},\vec{S}\,)$. Thus, we have shown that after the non--unitary
transformation of the Dirac fermion wave function $\psi \to (\sqrt{-
  \tilde{g}}\,\tilde{e}^0_{\hat{0}})^{1/2}\psi'$ in the Jordan frame
with the metric tensor $\tilde{g}_{\mu\nu}(x)$ the non--Hermitian
(anti--Hermitian) torsion--fermion interactions, violating $T$ and
$CP$ invariance, are conformal invariant. This agrees well with the
results, obtained by Silenko in Ref.\cite{Silenko2015}.

\subsection{The argument in behalf of observability of
    anti--Hermitian torsion--fermion interactions}

Now we would like to discuss a possible observability of the
anti--Hermitian torsion--fermion interactions, violating $CP$ and $T$
invariance. For the coordinate system, rotating with an angular
velocity $\vec{\omega}$, the vector $\vec{K}$ is equal to $\vec{K} = -
(\vec{\omega} \times \vec{r}\,)$
\cite{Hehl1990,Obukhov2009,Obukhov2011}. The effective low--energy
potential Eq.(\ref{eq:18}) is equal to
\begin{eqnarray}\label{eq:47}
\hspace{-0.3in}&&\tilde{\Phi}_{\rm eff}(t,\vec{r},\vec{S}\,) = \Phi^{(1)}_{\rm
  eff}(t,\vec{r},\vec{S}\,)\Big|_{\vec{K} = - (\vec{\omega} \times
  \vec{r}\,)} + \Phi^{(2)}_{\rm
  eff}(t,\vec{r},\vec{S}\,)\Big|_{\vec{K} = - (\vec{\omega} \times
  \vec{r}\,)} + \Phi^{(3)}_{\rm
  eff}(t,\vec{r},\vec{S}\,)\Big|_{\vec{K} = - (\vec{\omega} \times
  \vec{r}\,)} + \Phi^{(4)}_{\rm
  eff}(t,\vec{r},\vec{S}\,)\Big|_{\vec{K} = - (\vec{\omega} \times
  \vec{r}\,)}.\nonumber\\
\hspace{-0.3in}&&
\end{eqnarray}
Here the effective low--energy potential $\Phi^{(1)}_{\rm
  eff}(t,\vec{r},\vec{S}\,)\Big|_{\vec{K} = - (\vec{\omega} \times
  \vec{r}\,)}$ is 
\begin{eqnarray}\label{eq:48}
\Phi^{(1)}_{\rm eff}(t,\vec{r},\vec{S}\,)\Big|_{\vec{K} = -
  (\vec{\omega} \times \vec{r}\,)} &=& m (U_+ - U_{\rm E}) -
\vec{\omega}\cdot \vec{L} - \vec{\omega}\cdot \vec{S} -
\frac{1}{2}\,\vec{S}\cdot \vec{\cal B} + \frac{1}{3}\,{\cal
  K}\,\vec{S}\cdot (\vec{\omega} \times \vec{r}\,)\nonumber\\ &-&
\frac{1}{2}\,\vec{S}\cdot \Big(\vec{\cal M} \times (\vec{\omega}\times
\vec{r}\,)\Big) - \frac{1}{2}\,S_j\,\epsilon^{jk\ell}\,{\cal M}_{k\ell
  a}\,\epsilon^{abc}\omega_b x_c.
\end{eqnarray}
where $(\vec{\cal M}\,)_k = - {\cal M}_{00k}$ \cite{Ivanov2015b}. In
Eq.(\ref{eq:48}) the first term $m\,(U_+ - U_{\rm E})$ describes the
chameleon--matter interaction \cite{Brax2011,Ivanov2013}, whereas the
terms $ - \vec{\omega}\cdot \vec{L}$ and $ - \vec{\omega}\cdot
\vec{S}$, where $\vec{L} = - \vec{r}\times i\,\vec{\nabla}$ is the
orbital momentum operator of slow fermions (neutrons), agree well with
the results, obtained by Hehl and Ni \cite{Hehl1990}. The interactions
$- \vec{\omega}\cdot \vec{L}$ and $ - \vec{\omega}\cdot \vec{S}$ were
investigated and observed in the experiments by Werner, Staudenmann,
and Colella \cite{Werner1979}, by Atwood {\it et al.}
\cite{Atwood1984} and by Mashhoon \cite{Mashhoon1988}. Since the
effective low--energy potentials $\Phi^{(n)}_{\rm
  eff}(t,\vec{r},\vec{S}\,)\Big|_{\vec{K} = - (\vec{\omega} \times
  \vec{r}\,)}$ for $n = 2,3,4$, are calculated from the general
effective low--energy potential Eq.(\ref{eq:6}) as well as the
potential $\Phi^{(1)}_{\rm eff}(t,\vec{r},\vec{S}\,)\Big|_{\vec{K} = -
  (\vec{\omega} \times \vec{r}\,)}$, the observability of the
interactions $- \vec{\omega}\cdot (\vec{L} + \vec{S}\,)$ supports in
principle an observability of the torsion--fermion interactions in the
effective low--energy potential Eq.(\ref{eq:48}), the Hermitian
interactions $\Phi^{(2)}_{\rm eff}(t,\vec{r},\vec{S}\,)\Big|_{\vec{K}
  = - (\vec{\omega} \times \vec{r}\,)}$ and $\Phi^{(3)}_{\rm
  eff}(t,\vec{r},\vec{S}\,)\Big|_{\vec{K} = - (\vec{\omega} \times
  \vec{r}\,)}$ and, correspondingly, the anti--Hermitian interaction
$\Phi^{(4)}_{\rm eff}(t,\vec{r},\vec{S}\,)\Big|_{\vec{K} = -
  (\vec{\omega} \times \vec{r}\,)}$, respectively.

\section{Conclusion}
\label{sec:conclusion}

We have derived the operator of the angular velocity of the neutron
spin precession in the Einstein-Cartan gravity with torsion and
chameleon fields. For the calculation of such an operator we have used
the most general effective low--energy potential for slow Dirac
fermions, coupled to gravitational, chameleon and torsion fields to
order $1/m$, where $m$ is the fermion mass \cite{Ivanov2015b}. In
order to adapt such an effective low--energy potential to the
experimental analysis of the contributions of
fermion--gravitational,--chameleon and --torsion interactions we have
linearised it with respect to gravitational, chameleon and torsion
fields. Such a linearisation we have carried out in the curved
spacetime with the Schwarzschild metric, taken in the approximation of
the weak gravitational field and modified by a rotation with an
angular velocity $\vec{\omega}$ and the chameleon field. We have shown
that in the curved spacetime with such a modified metric torsion
scalar, pseudoscalar, axial--vector and tensor degrees of freedom
couple to slow neutrons through minimal torsion--fermion couplings
\cite{Ivanov2015b}. The obtained linearised effective low--energy
potential Eq.(\ref{eq:18}) is the generalization of the effective
low--energy potential, derived in \cite{Ivanov2015a}.

An important peculiarity of the linearised effective low--energy
potential Eq.(\ref{eq:18}) is the appearance of the anti--Hermitian
part. Such a part of the effective low--energy potential comes from
the operator $G_{\hat{j}}$ and proportional to the vector $K^{\hat{j}}
= - (\vec{\omega}\times \vec{r}\,)^{\hat{j}}$, related to a rotation
of a coordinate system.  A possible violation of Hermiticity of a
low--energy potential for slow fermions, coupled to gravitational,
chameleon and torsion fields in terms of the vierbein fields
Eq.(\ref{eq:4}), caused by the metric Eq.(\ref{eq:3}), might be
expected because of a non--invariance of the metric tensor
Eq.(\ref{eq:3}) and, correspondingly, Eq.(\ref{eq:13}) in a rotating
coordinate system with respect to time reversal transformation, i.e. a
non--invariance under $t \to - t$ transformation. We have found that
the anti--Hermitian part contains the terms proportional to vector
$K^{\hat{j}}$ and dependent on the torsion scalar ${\cal E}_{\hat{0}}$
and space--space--time ${\cal M}_{\hat{j}\hat{k}\hat{0}}$ and
time--space--space ${\cal M}_{\hat{0}\hat{k}\hat{j}}$ tensor degrees
of freedom. For the confirmation of the correctness of such an
anti--Hermitian part we have pointed out that the contribution of the
operator $G_{\hat{j}}$ is important for the correct derivation of the
effective low--energy potential for slow fermions (neutrons), coupled
to gravitational and chameleon fields. Another argument on behalf of
the correctness of such an anti--Hermitian part is the cancellation of
the torsion vector components in the $K_{\hat{j}}$--independent part
of the operator $G_{\hat{j}}$. Such a cancellation agrees well with
the results, obtained by Kostelecky \cite{Kostelecky2004} (see also
\cite{Ivanov2015a}). In the Appendix we have given a detailed
calculation of the operator $G_{\hat{j}}$ to linear order in the
gravitational, chameleon and torsion field approximation.

It is obvious that an anti--Hermitian part $\Phi^{(4)}_{\rm eff}(t,
\vec{r}, \vec{S}\,)$ of the effective low--energy potential $\Phi_{\rm
  eff}(t, \vec{r}, \vec{S}\,)$ violates ${\rm T}$ (or time reversal)
invariance. Since the effective low--energy potential $\Phi_{\rm
  eff}(t, \vec{r},\vec{S}\,)$ is invariant under ${\rm CPT}$
transformation, the anti--Hermitian part $\Phi^{(4)}_{\rm eff}(t,
\vec{r}, \vec{S}\,)$ violates also ${\rm CP}$ invariance,
i.e. invariance under Charge--Parity transformation. A violation of
${\rm CP}$ and ${\rm T}$ invariance in the spacetime with the
asymmetric Kerr metric, which is analogous to a spacetime in the
coordinate system rotating with an angular velocity
\cite{Obukhov2009}, has been discussed by Hadley \cite{Hadley2011}. We
would like also to mention that violation of parity and time reversal
invariance in the spin--rotation interactions has been discussed by
Papini \cite{Papini2002} and Scolarici and Solombrino
\cite{Scolarici2002} in the model with a modified Mashhoon's potential
of the spin--rotation coupling \cite{Mashhoon1988}.

Finally we would like to discuss the results, obtained in section
\ref{sec:obukhov}. As has been shown in
\cite{Obukhov2014}--\cite{Ivanov2015c} the well--known
non--Hermiticity of the Dirac Hamilton operator for relativistic
fermions, moving in the curved spacetime with an arbitrary metric
$\tilde{g}_{\mu\nu}(x)$ or in an arbitrary gravitational field, can be
removed by a non--unitary transformation $\psi \to (\sqrt{-
  \tilde{g}}\,\tilde{e}^{\hat{0}}_0)^{1/2} \psi'$, where $\tilde{g}$
is the determinant of the metric tensor $\tilde{g}_{\mu\nu}(x)$ and
$\tilde{e}^{\hat{0}}_0$ is the vierbein field. Such a property of the
Dirac Hamilton operator is also retained in the curved spacetime with
an arbitrary metric, torsion and chameleon field
\cite{Ivanov2015a,Ivanov2015b}. In section \ref{sec:obukhov} we have
shown that i) the anti--Hermitian Hamilton operator of
torsion--fermion interactions $\delta {\rm H}'_{\rm \bar{h}}$ in the
Dirac Hamilton operator ${\rm H}' = \gamma^{\hat{0}} m -
i\,\gamma^{\hat{0}}\vec{\gamma}\cdot \vec{\nabla} + \delta {\rm
  H}'_{\rm h} + \delta {\rm H}'_{\rm \bar{h}}$, obtained by means of
the non--unitary transformation of the Dirac fermion wave functions
$\psi \to (\sqrt{- \tilde{g}}\,\tilde{e}^{\hat{0}}_0)^{1/2} \psi'$,
cannot be removed by any additional non--unitary transformations and
ii) the existence of these anti--Hermitian torsion--fermion
interactions is fully caused by the properties of the curved
spacetimes with rotation, described by the metric Eq.(\ref{eq:3}).

We have also pointed out that our analysis of the Dirac Hamilton
operator and a derivation of the anti--Hermitian torsion--fermion
interactions are not related to the analysis of the non--Hermiticity
of the Dirac Hamilton operator by means of the $\eta$--representation
\cite{Gorbatenko2010}, requiring pseudo--Hermitian quantum mechanics
for a description of a dynamics of Dirac fermions in curved spacetimes
\cite{Gorbatenko2010,Parker1980,Bender2007}.

Then, we have discussed conformal invariance of the anti--Hermitian
torsion--fermion interactions. As has been shown by Silenko
\cite{Silenko2013}, quantum field theories of massless particles,
coupled to arbitrary gravitational fields or moving in curved
spacetimes with arbitrary metrics, are conformal invariant under
conformal transformation $\tilde{g}_{\mu\nu} \to O^2
\check{g}_{\mu\nu}$, where $O$ is a conformal factor.  The requirement
of conformal invariance of quantum field theories of massive particles
in curved spacetimes with arbitrary metrics can be fulfilled if and
only if particle masses are changed by the conformal factor as follows
$m \to O^{-1} \check{m}$ \cite{Silenko2015} (see also
\cite{Brans1961,Dicke1962}). We have shown that under the condition $m
\to O^{-1} \check{m}$ the relativistic anti--Hermitian Hamilton
operator $\delta {\rm H}_{\rm \bar{h}}$ and the anti--Hermitian
effective low--energy potential $\Phi^{(4)}_{\rm eff}(t, \vec{r},
\vec{S}\,)$ are conformal invariant.

We have discussed also an observability of the anti--Hermitian
torsion--fermion interactions. As a result, we may argue that the
anti--Hermitian torsion--fermion interactions can be in principle
observable. First, an observability of the obtained anti--Hermitian
torsion--fermion interactions is supported by their derivation,
carried out on the same footing as the effective low--energy
potentials $\Phi^{(1)}_{\rm eff}(t, \vec{r}, \vec{S}\,)$ and
$\Phi^{(1)}_{\rm eff}(t, \vec{r}, \vec{S}\,)$, which have been derived
earlier in \cite{Ivanov2015b,Hehl1990,Ivanov2014,Ivanov2015a}. Second,
an observability of the anti--Hermitian torsion--fermion interactions
Eq.(\ref{eq:22}) and Eq.(\ref{eq:38}) is supported experimentally as
follows. In the effective low--energy potential $\Phi^{(1)}_{\rm
  eff}(t, \vec{r}, \vec{S}\,)$ the interactions
\begin{eqnarray}\label{eq:49}
\delta \Phi^{(1)}_{\rm eff}(t,\vec{r},\vec{S}\,)\Big|_{\vec{K} = -
  (\vec{\omega} \times \vec{r}\,)} =  -
\vec{\omega}\cdot \vec{L} - \vec{\omega}\cdot \vec{S},
\end{eqnarray}
derived also by Hehl and Ni \cite{Hehl1990}, have been investigated
experimentally by Werner, Staudenmann, and Colella \cite{Werner1979},
by Atwood {\it et al.}  \cite{Atwood1984} and by Mashhoon
\cite{Mashhoon1988}. This makes reliable in principle an observability
of the anti--Hermitian torsion--fermion interactions, described by the
relativistic anti--Hermitian Hamilton operator Eq.(\ref{eq:38}) and
the anti--Hermitian effective low--energy potential Eq.(\ref{eq:22}).

The analysis of reliability and observability of the obtained
anti--Hermitian torsion--fermion interactions makes meaningful the
assumption that in the rotating Universe and galaxies \cite{Gamow1946}
(see also \cite{Obukhov1997}) the torsion--fermion interaction $\delta
{\rm H}_{\rm \bar{h}}$ in Eq.(\ref{eq:38}) as well as the low--energy
effective potential $\Phi^{(4)}_{\rm eff}(t, \vec{r}, \vec{S}\,)$ in
Eq.(\ref{eq:22}) might be an origin of i) violation of $CP$ and $T$
invariance in the Universe and ii) of baryon asymmetry \cite{PDG2014}.

\section{Acknowledgements}

We are grateful to Hartmut Abele for stimulating discussions. This
work was supported by the Austrian ``Fonds zur F\"orderung der
Wissenschaftlichen Forschung'' (FWF) under the contract I689-N16.

\section{Appendix A: Detailed calculation of the operator $G_{\hat{j}}$}
\renewcommand{\theequation}{A-\arabic{equation}}
\setcounter{equation}{0}

In the Appendix we give a detailed calculation of the operator
$G_{\hat{j}}$. According to Eq.(\ref{eq:7}) it is defined by
\begin{eqnarray}\label{eq:A.1}
 G_{\hat{j}}
&=& \frac{1}{2}\, \tilde{e}^{\hat{0}}_0(x)\,\Big({\tilde{\cal
    T}^{\alpha}\,\!\!}_{\alpha\ell}(x) \,\tilde{e}^{\ell}_{\hat{j}}(x)
+ \tilde{\omega}_{0\hat{j}\hat{0}}(x)\, \tilde{e}^0_{\hat{0}}(x) +
\tilde{\omega}_{\ell\hat{j}\hat{0}}(x)\, \tilde{e}^{\ell}_{\hat{0}}(x)
+ \tilde{\omega}_{\ell\hat{j}\hat{k}}(x)\,
\tilde{e}^{\ell}_{\hat{\ell}}(x)\,\eta^{\hat{\ell}\hat{k}}\Big)\nonumber\\ &+&
\frac{1}{2}\,(\tilde{e}^{\hat{0}}_0(x))^2\,\tilde{e}^j_{\hat{j}}(x)
\,\frac{1}{\sqrt{- \tilde{g}(x)}}\,\frac{\partial }{\partial
  x^j}\Big(\sqrt{-
  \tilde{g}(x)}\,\tilde{e}^0_{\hat{0}}(x)\Big).
\end{eqnarray}
Using $\sqrt{-\tilde{g}} = 1 + U_+ - 3 U_-$ \cite{Ivanov2015b} and the
vierbein fields Eq.(\ref{eq:13}) we transcribe the r.h.s. of
Eq.(\ref{eq:A.1}) into the form
\begin{eqnarray}\label{eq:A.2}
G_{\hat{j}} &=&
\frac{1}{2}\, {\tilde{\cal T}^{\alpha}\,\!\!}_{\alpha \hat{j}}(x) +
\frac{1}{2}\,\tilde{\omega}_{0\hat{j}\hat{0}}(x) +
\frac{1}{2}\,\tilde{\omega}_{\ell\hat{j}\hat{0}}(x)\,K^{\ell} +
\frac{1}{2}\,\tilde{\omega}_{\ell\hat{j}\hat{k}}(x)\, \eta^{\ell\hat{k}}  -
\frac{3}{2}\,\frac{\partial U_+}{\partial x^{\hat{j}}},
\end{eqnarray}
where we have made the following replacements $(1 + U_+ + U_-)
{\tilde{\cal T}^{\alpha}\,\!\!}_{\alpha \hat{j}}(x) \to {\tilde{\cal
    T}^{\alpha}\,\!\!}_{\alpha \hat{j}}(x)$ and $ 1 + U_+ + U_-)
    \tilde{\omega}_{\ell\hat{j}\hat{k}}(x)\, \eta^{\ell\hat{k}} \to
    \tilde{\omega}_{\ell\hat{j}\hat{k}}(x)\, \eta^{\ell\hat{k}}$ and
    calculated
\begin{eqnarray}\label{eq:A.3}
\frac{1}{2}\,(\tilde{e}^{\hat{0}}_0(x))^2\,\tilde{e}^j_{\hat{j}}(x)
\,\frac{1}{\sqrt{- \tilde{g}(x)}}\,\frac{\partial }{\partial
  x^j}\Big(\sqrt{- \tilde{g}(x)}\,\tilde{e}^0_{\hat{0}}(x)\Big) = -
\frac{3}{2}\,\frac{\partial U_+}{\partial x^{\hat{j}}},
\end{eqnarray}
keeping only the linear order contributions of the gravitational,
chameleon and torsion fields. Since we keep the contributions to order
$O(K^{\hat{j}})$, we get
\begin{eqnarray}\label{eq:A.4}
\frac{1}{2}\,{\tilde{\cal T}^{\alpha}\!}_{\alpha \hat{j}}(x) &=&
\frac{1}{2}\, {\cal T}_{\hat{0}\hat{0}\hat{j}} + \frac{1}{2}\,{\cal
  T}_{\hat{\ell}\hat{k} \hat{j}}\,\eta^{\hat{\ell}
  \hat{k}},\nonumber\\ \frac{1}{2}\,\tilde{\omega}_{0\hat{j}\hat{0}}(x) &=&
\frac{1}{2}\,K_{\hat{j}}\,\frac{\partial U_-}{\partial t} -
\frac{1}{2}\,\frac{\partial U_+}{\partial x^{\hat{j}}} +
\frac{1}{2}\,{\cal K}_{\hat{j}\hat{0}\hat{0}} + \frac{1}{2}\,{\cal
  K}_{\hat{j}\hat{0}\hat{\ell}}\,K^{\hat{\ell}},\nonumber\\ \frac{1}{2}\,
\tilde{\omega}_{\ell\hat{j}\hat{0}}(x)\, K^{\ell} &=& -
\frac{1}{2}\,K_{\hat{j}}\,\frac{\partial U_-}{\partial t} +
\frac{1}{2}\,{\cal
  K}_{\hat{j}\hat{\ell}\hat{0}}\,K^{\hat{\ell}},\nonumber\\ \frac{1}{2}\,
\tilde{\omega}_{\ell\hat{j}\hat{k}}(x)\, \eta^{\ell\hat{k}} &=&
\frac{\partial U_-}{\partial x^{\hat{j}}} + \frac{1}{2}\,{\cal
  K}_{ \hat{j}\hat{\ell}\hat{k}}\,\eta^{\hat{\ell}
 \hat{k}} - \frac{1}{2}\,K_{\hat{j}}\,{\cal
  K}_{ \hat{0}\hat{\ell}\hat{k}}\,\eta^{\hat{\ell}\hat{k}}.
\end{eqnarray}
Plugging Eq.(\ref{eq:A.4}) into Eq.(\ref{eq:A.2}) we derive the
operator $G_{\hat{j}}$ in the following form
\begin{eqnarray}\label{eq:A.5}
 G_{\hat{j}} =  - \frac{1}{2}\,\frac{\partial}{\partial
   x^{\hat{j}}}(U_+ + U_-) + \frac{1}{2}\,\Big({\cal
   T}_{\hat{0}\hat{0}\hat{j}} +{\cal T}_{\hat{\ell}\hat{k}
   \hat{j}}\,\eta^{\hat{\ell} \hat{k}} + {\cal
   K}_{\hat{j}\hat{0}\hat{0}} + {\cal K}_{
   \hat{j}\hat{\ell}\hat{k}}\,\eta^{\hat{\ell} \hat{k}}\Big) +
 \frac{1}{2}\,\Big(({\cal K}_{\hat{j}\hat{0}\hat{\ell}} + {\cal
   K}_{\hat{j}\hat{\ell}\hat{0}})\,K^{\hat{\ell}} - K_{\hat{j}}\,{\cal
   K}_{ \hat{0}\hat{\ell}\hat{k}}\,\eta^{\hat{\ell}\hat{k}}\Big).
\end{eqnarray}
Using the properties of the contorsion tensor \cite{Ivanov2015b} and
the irreducible representation of torsion Eq.(\ref{eq:15}) we
transcribe Eq.(\ref{eq:A.5}) into Eq.(\ref{eq:26}) (see also
Eq.(\ref{eq:14})).

\end{document}